\documentclass[12pt]{spieman}  % 12pt font required by SPIE;
\usepackage{amsmath,amsfonts,amssymb}
\usepackage{graphicx}
\usepackage{setspace}
\usepackage{tocloft}
\usepackage{lineno}
\title{Telescope control software and proto-model siderostat for the SDSS-V Local Volume Mapper}

\author[a]{Hojae Ahn}
\author[b]{Florian Briegel}
\author[a]{Jimin Han}
\author[a]{Mingyu Jeon}
\author[b]{Thomas M. Herbst}
\author[a]{Sumin Lee}
\author[c]{Woojin Park}
\author[a]{Sunwoo Lee}
\author[d]{Inhwan Jung}
\author[a]{Tae-Geun Ji}
\author[a]{Changgon Kim}
\author[e]{Geon Hee Kim}
\author[b]{Wolfgang Gaessler}
\author[b]{Markus Kuhlberg}
\author[f]{Hyun Chul Park}
\author[a,*]{Soojong Pak}
\author[g]{Nicholas P. Konidaris}
\author[h]{Niv Drory}
\author[i]{Jos\'{e} R. S\'{a}nchez-Gallego}
\author[h]{Cynthia S. Froning}
\author[g]{Solange Ramirez}
\author[g]{Juna A. Kollmeier}
\affil[a]{Kyung Hee University, School of Space Research, 1732 Deogyeong-daero, Giheung-gu, Yongin-si, Gyunggi-do, Republic of Korea, 17104}
\affil[b]{Max Planck Institute for Astronomy, K\"{o}nigstuhl 17, Heidelberg, Germany, 69117}
\affil[c]{Korea Astronomy and Space Science Institute, 776 Daedeokdae-ro, Yuseong-gu, Daejeon, Republic of Korea, 34055}
\affil[d]{Kyung Hee University, Graduate school of Physics Education, 24 Kyungheedae-ro, Dongdaemun-gu, Seoul, Republic of Korea, 02453}
\affil[e]{Hanbat University, Department of Mechanical and Material Convergence System Engineering, 125 Dongseo-daero, Yuseong-gu, Daejeon, Republic of Korea, 34158}
\affil[f]{Konkuk University, 120, Neungdong-ro, Gwangjin-gu, Seoul, Republic of Korea, 05029}
\affil[g]{Carnegie Institution for Science, 813 Santa Barbara Street, Pasadena, CA, USA, 91101}
\affil[h]{The University of Texas at Austin, Department of Astronomy, 2515 Speedway, Stop C1400, Austin, TX, USA, 78712-1205}
\affil[i]{University of Washington, Department of Astronomy, 3910 15th Ave NE, Seattle WA, USA, 98195}

\cftpagenumbersoff{figure}
\cftpagenumbersoff{table} 
\begin{document} 
\maketitle

\begin{abstract}
The fifth Sloan Digital Sky Survey (SDSS-V) Local Volume Mapper (LVM) is a wide-field integral field unit (IFU) survey that uses an array of four 160 mm fixed telescopes with siderostats to minimize the number of moving parts. Individual telescope observes the science field or calibration field independently and is synchronized with the science exposure. We developed the LVM Acquisition and Guiding Package (LVMAGP) optimized telescope control software program for LVM observations, which can simultaneously control four focusers, three K-mirrors, one fiber selector, four mounts (siderostats), and seven guide cameras. This software is built on a hierarchical architecture and the SDSS framework and provides three key sequences: autofocus, field acquisition, and autoguide. We designed and fabricated a proto-model siderostat to test the telescope pointing model and LVMAGP software. The mirrors of the proto-model were designed as an isogrid open-back type, which reduced the weight by 46\% and enabled reaching thermal equilibrium quickly. Additionally, deflection due to bolting torque, self-gravity, and thermal deformation was simulated, and the maximum scatter of the pointing model induced by the tilt of optomechanics was predicted to be $4'.4$, which can be compensated for by the field acquisition sequence. We performed a real sky test of LVMAGP with the proto-model siderostat and obtained field acquisition and autoguide accuracies of $0''.38$ and $1''.5$, respectively. It met all requirements except for the autoguide specification, which will be resolved by more precise alignment among the hardware components at Las Campanas Observatory.
\end{abstract}

% Include a list of up to six keywords after the abstract
\keywords{astronomy, coding, telescopes, pointing, tracking, architectures}

% Include email contact information for corresponding author
{\noindent \footnotesize\textbf{*}Soojong Pak, \linkable{soojong@khu.ac.kr}}

\begin{spacing}{1.2}   % use double spacing for rest of manuscript

\section{Introduction}
\label{sec:intro} 
The Sloan Digital Sky Survey (SDSS) has been providing all-sky photometric and spectroscopic data on numerous stars, quasars, and galaxies since 2000\cite{York+00, Frieman+08, Yanny+09, Eisenstein+11}. So far, SDSS observations have been conducted using two 2.5 m telescopes: Sloan Foundation Telescope at Apache Point Observatory (APO) for the northern hemisphere and Ir\'{e}n\'{e}e du Pont Telescope at Las Campanas Observatory (LCO) for the southern hemisphere\cite{Gunn+06,Bowen+73}. The fifth Sloan Digital Sky Survey (SDSS-V) is the first panoptic and multi-epoch spectral survey to continue the strong SDSS legacy of innovative data production\cite{Kollmeier+19}. As part of the SDSS-V, the Local Volume Mapper (LVM) aims to conduct a wide integral-field spectroscopic survey to probe the physical state of ionized interstellar medium (ISM) in the Milky Way, Magellanic Clouds, and local group galaxies. This survey spans 2,500 square degrees with sub-parsec scale spatial resolution and covers the optical wavelength range (360--980 nm) with a spectral resolution of R$\sim$4,000 at the H$\alpha$ wavelength\cite{Kollmeier+19, Konidaris+20}. A similar survey using an integral field unit (IFU) was performed by the SDSS-IV Mapping Nearby Galaxies at APO (MaNGA)\cite{Bundy+15}. The significant difference between the MaNGA and LVM platforms is the angular size of the target; thus, MaNGA can observe distant galaxies while the LVM can observe ISM in the Milky Way and local group galaxies. As a result, the LVM telescope should cover a much wider field of view than the MaNGA, given that the two 2.5 m telescopes have focal lengths that are too long for the LVM.

To address these unique requirements, we designed specialized hardware and software for the LVM, referred to as the LVM instrument (LVMi), which will be installed at the LCO. The LVMi consists of four subsystems, namely, enclosure, telescope, fiber, and spectrograph, with each subsystem having its own instrument control software (ICS)\cite{Konidaris+20}. The LVMi telescope subsystem employs four 160 mm telescopes for science data acquisition (sci telescope), sky subtraction (skye and skyw telescopes), and spectrophotometric calibration (spec telescope). Unlike conventional telescopes, these telescopes are configured with siderostats to stabilize the system by minimizing moving parts\cite{Herbst+20}. Thus, the optical elements of the LVM lie on the optical bench without movement, regardless of pointing. The siderostat is mounted with altitude-altitude (alt-alt) configurations to avoid the zenith blind spot\cite{Richardson+90, Riva+10}. Notably, each telescope does not have its own CCD. The light from the four telescope units is gathered through a fiber cabling system and distributed to three spectrographs\cite{Konidaris+20}.

To ensure that the LVMi configuration is operable, all telescopes must be synchronized by CCD exposure, which is the ultimate function of the ICS in the telescope subsystem. Asynchronous programming facilitates real-time device monitoring and control with reduced response delays compared to that observed with synchronous timers and loops\cite{Durech+18, Reefe+22}. In addition, code reuse,\cite{Dipper+04, Park+22TCS, Kim+23} containerization,\cite{Bento+22} and Agile development\cite{Ji+20,Baxter+22} have become important for improving code quality and productivity for ICS development. By adopting these components, we developed the LVM Acquisition and Guiding Package (LVMAGP) software program that supports robust control of instruments under the telescope subsystem, including four custom telescope units with four mounts (siderostats), four focusers, seven guide cameras, three K-mirrors, and one fiber selector.

Hardware and software in LVMi were parallelly developed. As a result, we made a proto-model siderostat for the software test during the development phase. This study details the development, functionality, and performance of the LVMAGP and proto-model siderostat. In Sec.~\ref{sec:LVMAGP Requirements and Software Architectures}, the system hierarchy, including the software and hardware, and the system requirements and class diagram of LVMAGP are listed. In Sec.~\ref{sec:SDSS Framework}, the implementation of environmental requirements within the SDSS framework is discussed. In Sec.~\ref{sec:LVMAGP Sequence Development}, the implementation of the functional requirements through a sequence diagram with algorithms is presented. In Sec.~\ref{sec:Proto-model Siderostat}, the design and analysis of the proto-model siderostat for software testing are presented. In Sec.~\ref{sec:performance}, the setup and results of performance tests are discussed. Finally, in Sec.~\ref{sec:conclusion}, a summary of the study is provided.

\section{LVMAGP Requirements and Architecture} \label{sec:LVMAGP Requirements and Software Architectures}

\subsection{LVMi System Architecture} \label{subsec:LVMi System Architecture}
\begin{figure} [ht]
   \begin{center}
   \begin{tabular}{c} 
   \includegraphics[width=\textwidth]{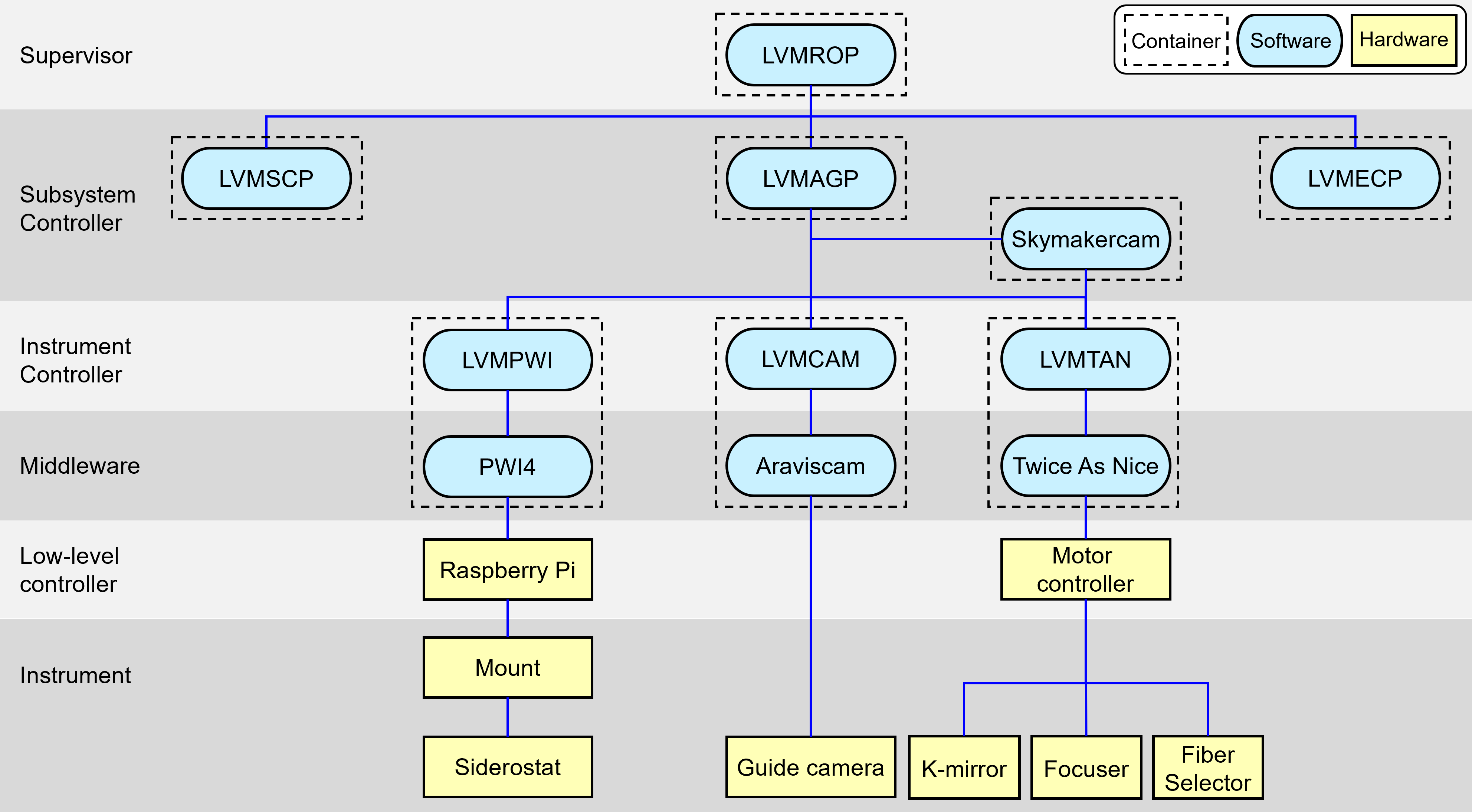}
   \end{tabular}
   \end{center}
   \caption[System architecture of the LVMi telescope subsystem] 
   { \label{fig:architecture} 
System architecture of the LVMi telescope subsystem. Lower-level components under LVMSCP and LVMECP are not presented.}
   \end{figure}

A hierarchical structure is well suited for managing complex systems composed of both software and hardware\cite{Ravanmehr+14,Ji+20}. The LVMi also has a hierarchical and modular structure for efficient instrument control and software development and contains a software to ensure independence and ease of maintenance (Fig.~\ref{fig:architecture}). At the top, the LVM Robotic Operation Package (LVMROP) supervises all observational operations, safety control, and target tiling. Under the LVMROP, subsystem controllers are arranged, including the LVMAGP (telescope subsystem), LVMSCP\cite{Kim+23, lvmscp} (spectrograph subsystem), and LVMECP\cite{lvmecp} (enclosure subsystem). To perform observations, LVMROP allocates tasks to the relevant subsystem controllers, which in turn delegate tasks to the instrument controllers. Subsequently, the instrument controllers and instruments perform the work and report the results, which are relayed back to the LVMROP. Instrument controllers exclusively handle communication with instruments and subsystem controllers, whereas supervisor-level software is responsible for executing logical operations and sequences. Here, we describe the instruments and their controllers under LVMAGP in detail.

For the siderostat, we used a mount (L-350, PlaneWave Instrument Inc.) in a north-south alt-alt configuration by vertically mounting it to a wedge. The LVM PlaneWave Interface (LVMPWI) serves as a wrapper for PlaneWave Interface 4 (PWI4), a control software provided by the manufacturer, thereby enabling the telescope to slew to specified positions and track targets based on a pointing model, PointXP, also provided by the manufacturer\cite{lvmpwi,L350}. Using pointing data of multiple stars, this model is optimized to compensate for mechanical errors such as polar misalignment, non-orthogonality of the mount axes, or flexure of the optical tube. LVMPWI has a built-in simulator for development and testing purposes. The TwiceAsNice (TAN) control package (LVMTAN) manages custom motor controllers to adjust the focusers, compensate for field rotation with K-mirrors, and control the fiber selector\cite{lvmtan}. It has its own simulator and calculates the rotation of K-mirrors to follow the trajectory of the target. The LVM CAMera control package (LVMCAM) operates guide cameras (Blackfly S BFS-PGE-16S7M, Teledyne FLIR, LLC) and saves the images in FITS format\cite{lvmcam}. The guiding system has a pixel scale of $1''.0$/pixel, providing field of view of $27'\times18'$. Skymakercam, a simulator for LVMCAM, generates realistic simulated images by considering the source (star position and magnitude from Gaia Early Data Release 3\cite{Gaia+21}), telescope information (focal ratio, pixel scale, defocus, and position angle with K-mirror rotation), sensor parameters (bias, dark current, exposure time, gain, and read-out noise), and environmental factors (seeing and sky background).

\begin{figure} [ht]
    \begin{center}
    \begin{tabular}{c} 
    \includegraphics[width=\textwidth]{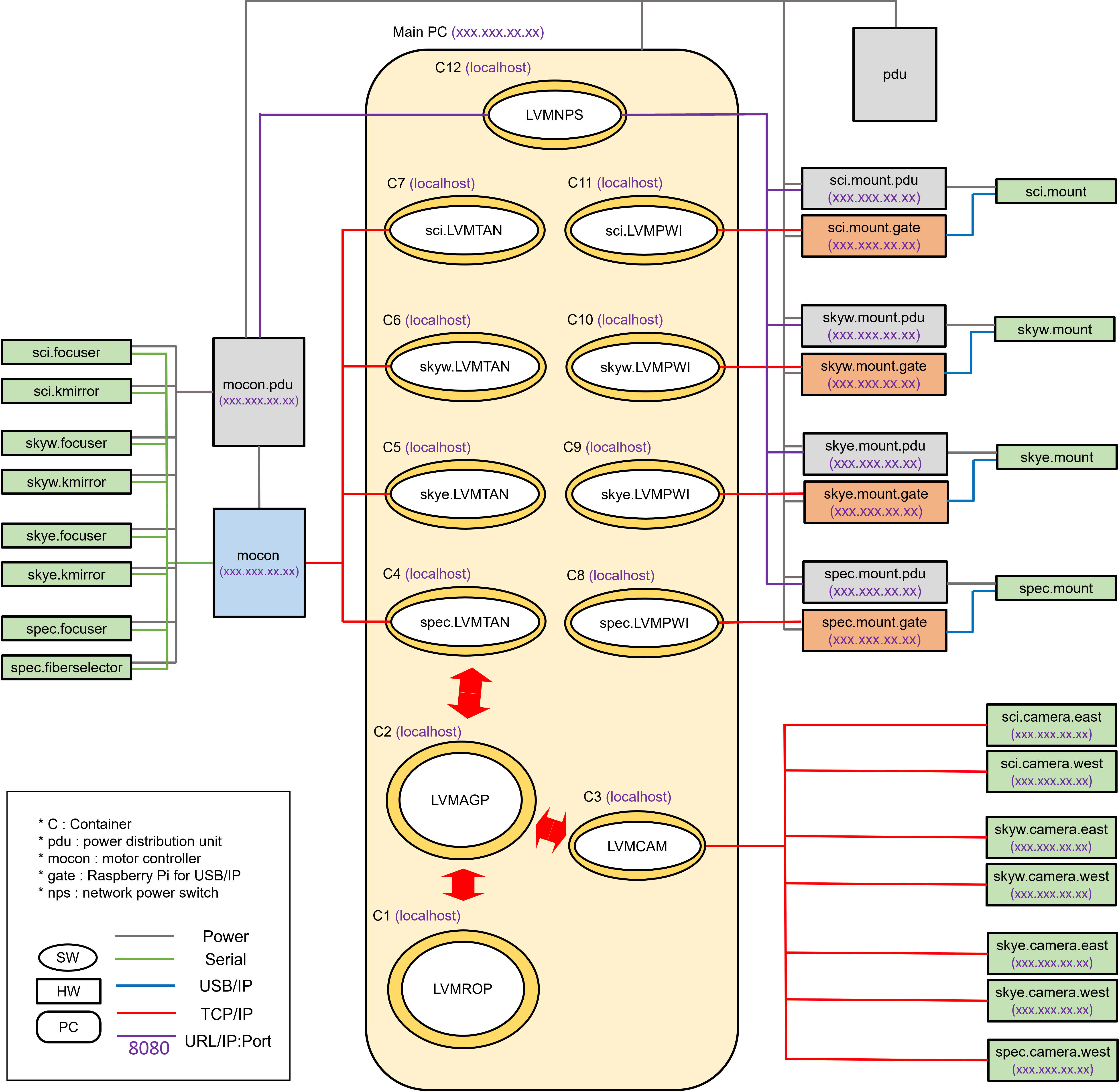}
    \end{tabular}
    \end{center}
    \caption[System architecture, including all individual instruments and ICS instances under the telescope subsystem] 
    { \label{fig:architecture_allinst} 
    System architecture, including all individual instruments and ICS instances under the telescope subsystem. Each system element is denoted in the format ``telescope unit.element." Connections between components are illustrated by lines, with the type of connection indicated by the color of line. The chrome yellow oval indicates the container where the software runs. LVMAGP communicates with controllers in containers C4 to C11, although the arrow only points to the spec.LVMTAN actor to simplify the figure.
    }
    \end{figure}

Figure~\ref{fig:architecture_allinst} illustrates a real situation involving multiple devices, including four telescope units: sci, skyw, skye, and spec. Each unit requires a distinct instance of the instrument controller operating within isolated containers. For LVMCAM, only one instance suffices because it accommodates multiple devices in one actor instance via the Araviscam module\cite{araviscam}. Figure~\ref{fig:architecture_allinst} also indicates the methods for hardware-software and software-software communications. All software communications are processed based on TCP/IP communication. However, the software-hardware communication method is not unified and may include USB/IP communication, URL/IP communication, or TCP/IP communication. Electric power is distributed by a power distribution unit (PDU) controlled by a network power switch (NPS). The instrument controller for the NPS is the LVMNPS, which is under the LVMSCP\cite{lvmnps}.

\subsection{System Requirements} \label{subsec:System Requirements}
The SDSS software must meet system specifications, which range from instrument requirements to the development environment. Table \ref{tab:sysreq} describes the key requirements for the LVMAGP, including the development and deployment environments (Reqs.~1--6), target instruments (Req.~7), key functions (Reqs.~8--10), and each key function specification (Reqs. 11--13). We refer to the key functions of LVMAGP as autofocus (Req.~8), field acquisition (Req.~9), and autoguide (Req.~10). Req.~7 implies that an asynchronous platform is required to enable the simultaneous operation of multiple instruments across the four telescope units.

To fulfill Req.~1, we developed LVMAGP using Python 3.7 on Linux. In terms of technical tools, PyCharm, an Integrated Development Environment (IDE), was used to facilitate an efficient coding process. Reqs.~2--6 were achieved by adopting the SDSS framework. To address Reqs.~7--13, we implemented three sequential commands corresponding to the key functionalities. To avoid reinventing the wheel (Reqs.~6 and 11), we intended to utilize the PWI4 software in LVMPWI. However, the mount and PWI4 were originally designed for alt-azimuth and equatorial configurations and have never been tested with a siderostat with the alt-alt configuration. Thus, we constructed a proto-model siderostat to verify the tracking capabilities of the mount and the applicability of the pointing model to the alt-alt configuration. This prototype also served to assess whether the LVMAGP met the functional requirements.

    \begin{table}[ht]
    \caption{Key system requirements for the LVMAGP} 
    \label{tab:sysreq}
    \begin{center}       
    \begin{tabular}{ll} 
    \hline
    \rule[-1ex]{0pt}{3.5ex}  Req. No. & Description\\
    \hline
    \rule[-1ex]{0pt}{3.5ex}  Req. 1. & All production software recommends Python 3.7+ on Linux.\\
    \hline
    \rule[-1ex]{0pt}{3.5ex}  Req. 2. & The version of LVMi software should be updated continuously and maintained at\\
    &\url{https://github.com/sdss}. \\
    \hline
    \rule[-1ex]{0pt}{3.5ex}  Req. 3. & LVMi software must be deployed in isolated environments or Docker.\\
    \hline
    \rule[-1ex]{0pt}{3.5ex}  Req. 4. & LVMi ICS functions should be called concurrently.\\
    \hline 
    \rule[-1ex]{0pt}{3.5ex}  Req. 5. &  Message protocol in the LVMi instrument control software (ICS) design should\\
    & follow the SDSS software framework AMQP/RabbitMQ. \\
    \hline
    \rule[-1ex]{0pt}{3.5ex}  Req. 6. & SDSS software recommends recycling resources, including code reuse.\\
    \hline
    \rule[-1ex]{0pt}{3.5ex}  Req. 7. &  LVMAGP should command the mounts (4 units), K-mirrors (3 units),\\
    & focusers (4 units), fiber selector (1 unit),\\
    & and guide cameras (7 units) individually and concurrently. \\
    \hline 
    \rule[-1ex]{0pt}{3.5ex}  Req. 8. &  LVMAGP will perform a focus sequence at the beginning of each observing night\\
    & and when the ambient temperature changes. \\
    \hline
    \rule[-1ex]{0pt}{3.5ex}  Req. 9. & LVMAGP should be able to point at any RA/Dec/PA or HA/Dec/PA,\\
    & with guiding on/off.\\
    \hline
    \rule[-1ex]{0pt}{3.5ex}  Req. 10. &  LVMAGP should be able to execute offsets from the given pointing either in\\
    & $\Delta$RA/$\Delta$Dec/$\Delta$PA (in arcsecond units) or $\Delta$x/$\Delta$y/$\Delta$PA (in pixel or arcsecond units\\
    & relative to the focal plane axes).\\
    \hline
    \rule[-1ex]{0pt}{3.5ex}  Req. 11. & Telescope pointing solution should be optimized to correct axis misalignment\\
    & and similar sources of pointing errors. \\
    \hline 
    \rule[-1ex]{0pt}{3.5ex}  Req. 12. &  Acquisition and guiding errors must be $<1$ arcsecond in RMS.\\
    \hline
    \rule[-1ex]{0pt}{3.5ex}  Req. 13. &  LVMAGP should perform autoguide using more than one star to track rotation\\
    & and scale and provide better statistics.\\
    \hline
    \end{tabular}
    \end{center}
    \end{table}

\subsection{Software Architecture} \label{subsec:Software Architecture}
\begin{figure} [ht]
   \begin{center}
   \begin{tabular}{c} 
   \includegraphics[width=\textwidth]{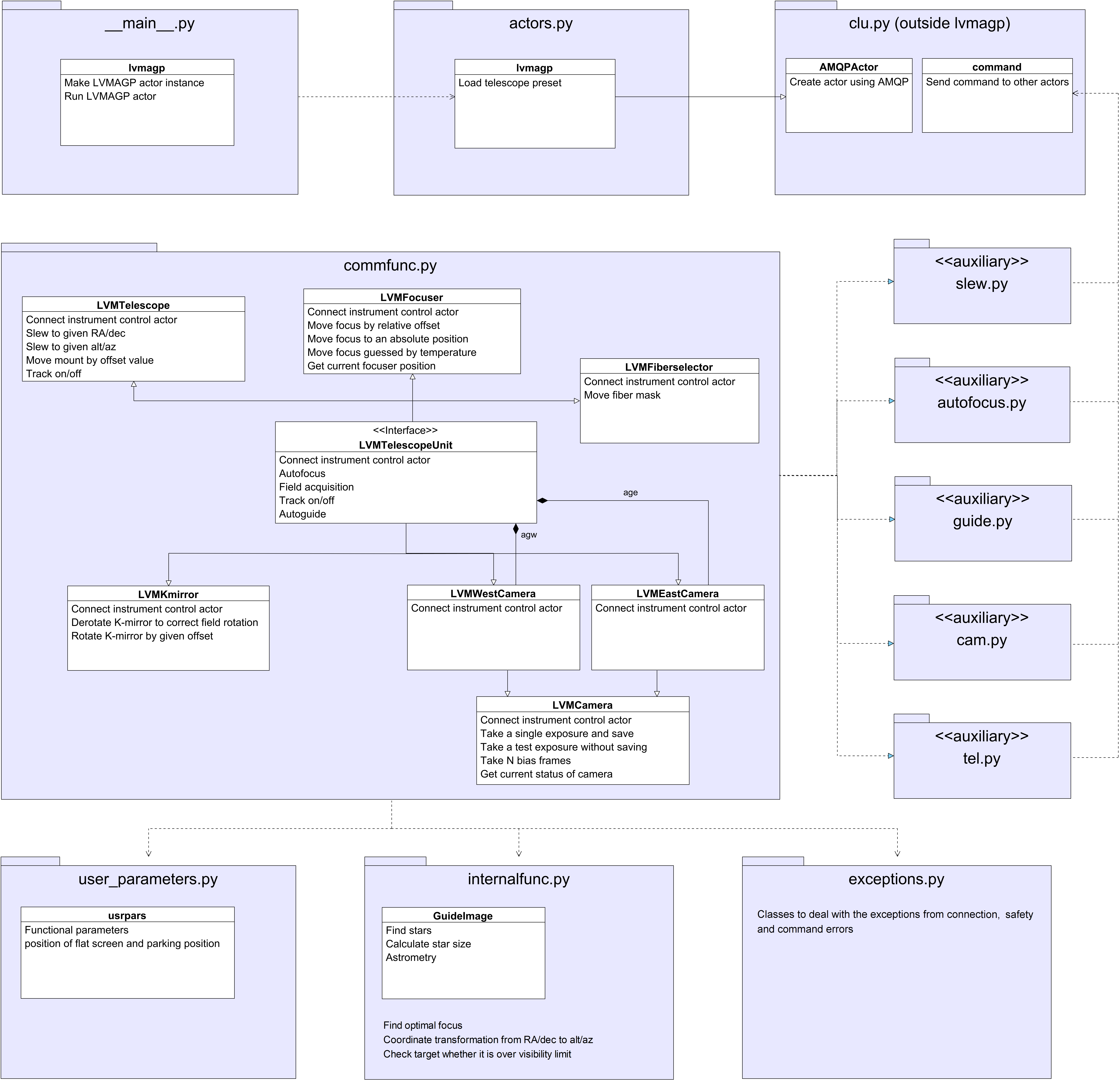}
   \end{tabular}
   \end{center}
   \caption[Class diagram of LVMAGP] 
   { \label{fig:software_architecture} 
Class diagram of LVMAGP. The blue file-like icons indicate Python files, and the white boxes indicate the classes in the Python file.}
   \end{figure}

Figure~\ref{fig:software_architecture} shows the relationship between Python files and the classes comprising LVMAGP. The \ttfamily \_\_main\_\_.py\rmfamily, \ttfamily actors.py\rmfamily, and \ttfamily clu.py\rmfamily\; files are associated with initiating the software and the message-passing server. The \ttfamily commfunc.py\rmfamily\; file defines the classes for communication with lower-level controllers. The \ttfamily user\_parameters.py\rmfamily\; file defines user parameters, such as the exposure time for autoguiding. The \ttfamily internalfunc.py\rmfamily\; file performs data calculations (e.g., current telescope position from astrometry and conversion between image and celestial coordinates), which are sent to other hardware and software. Finally, the \ttfamily exceptions.py\rmfamily\; file defines the exceptions based on connection, safety, and command failures.

LVMAGP can be utilized via two methods. The first method uses an API. The \ttfamily LVMTelescopeUnit\rmfamily\;\\
class acts as an interface between users and low-level classes. The key functionality of LVMAGP can be invoked by importing the \ttfamily LVMTelescopeUnit\rmfamily\; class. Because four telescope units (sci, skye, skyw, and spec) are included, four instances of \ttfamily LVMTelescopeUnit\rmfamily\; should be created, and these instances will independently control each instrument unit. This class inherits the low-level classes that communicate with the instrument controllers. Thus, LVMAGP satisfies Req.~7 obtaining access to control instruments through low-level classes. The \ttfamily LVMTelescope\rmfamily\; class communicates with LVMPWI. The \ttfamily LVMFocuser\rmfamily, \ttfamily LVMFiberselector\rmfamily, and \ttfamily LVMKmirror\rmfamily\; classes interact with LVMTAN. The \ttfamily LVMCamera\rmfamily\; class is connected to LVMCAM. The \ttfamily LVMEastCamera\rmfamily\; and \ttfamily LVMWestCamera\rmfamily\; classes are used to distinguish between the two guide cameras (east and west in Fig.~\ref{fig:architecture_allinst}) in one telescope unit, and they inherit the \ttfamily LVMCamera\rmfamily\; class without additional methods. 

The second method uses the LVMAGP command-line user interface (CLI) for direct user interactions. In this case, LVMAGP runs on a message-passing server (Codified Likeness Utility (CLU), described in Sec.~\ref{subsec:clu}). When a user inputs `lvmagp start' into the CLU terminal, \ttfamily main.py\rmfamily\; is executed, and LVMAGP is initiated by creating an actor through the CLU. Once operational, we can operate the telescope systems using predefined commands, such as slew, autofocus, guide, cam, and tel. The execution of these commands triggers the execution of a Python file whose name is the same as that of the command. This Python file is a wrapper of the methods in the \ttfamily LVMTelecopeUnit\rmfamily\; class; therefore, the methods in the \ttfamily LVMTelescopeUnit\rmfamily\; are executed.

\section{SDSS Framework} \label{sec:SDSS Framework}
The SDSS framework is a loose set of software, design methodologies and nomenclature that has become the standard for SDSS software development over the past 20 years. It also follows modern trends to improve code quality and productivity, as mentioned in Sec.~\ref{sec:intro}. In this section, we introduce the main factors that ensure that LVMAGP satisfies Reqs. 2--6.

\subsection{Actor and CLU} \label{subsec:clu}
The SDSS actor (hereafter actor) is a piece of software that sends or receives messages and performs actions. These actors serve as the building blocks for the modular hierarchy of the ICS. The software in the supervisor, subsystem controller, and instrument controller layers, shown in Fig.~\ref{fig:architecture}, are actors. Throughout the remainder of this study, LVMPWI, LVMCAM, and LVMTAN will be referred as low-level actors. For efficient and consistent development, the skeleton actor can be generated from a template\cite{python_template} using cookiecutter\cite{cookiecutter}.

For actor-to-actor communication, SDSS has adopted AMQP\cite{Vinoski06}/RabbitMQ\cite{RabbitMQ} as the preferred message-passing protocol and developed the Codified Likeness Utility (CLU), a protocol wrapper\cite{clu}. CLU provides a CLI built upon a Command Line Interface Creation Kit (CLICK)\cite{click}-enabled command parser. Therefore, actors are command line-based. Another characteristic of the CLU is that it runs asynchronously by importing asyncio\cite{asyncio}, a Python library used for asynchronous programming. Thus, the actor is also asynchronous. The actor manages the traffic of commands; that is, it runs the commands sequentially or concurrently as necessitated by the system's demands. CLUplus\cite{cluplus} enhances this framework by wrapping asynchronous actor commands into synchronous methods in a class named after the actor. This allowed the commands of one actor to be invoked from other Python software by importing the actor as a Python module, thereby demonstrating the actor's API-like functionality for external Python applications.

\subsection{Containerization} \label{subsec: Containerization}
Low-level actors control only a single device. However, there are diverse hardware types and distinct units within each type. To address this issue, actors were deployed in Docker\cite{Docker} or Podman\cite{Podman} containers to establish separate and isolated environments for each instance. This allows multiple instances of the same actor to be created; for example, one for a guide camera attached to a science telescope and another for a guide camera with a spectrophotometric telescope.

Another advantage of containerized actors is that containerization is a user-friendly deployment method. The container is created from the image, which is an immutable template containing all files, environments, and instructions required to execute the software. Thus, the containerization obviates the need for manual installation of Python packages or third-party software and mitigates issues related to Python package dependencies or operating system compatibility. Users can simply download and build the deployed image, and an additional installation is not required to use the actor. Consequently, the actor can operate in a consistent and stable environment on different PCs.

While the software is containerized for deployment, the actual development occurs outside the container. Tools such as Virtualenv and Poetry are employed to effectively manage Python package dependencies during the development phase. The Git repository plays pivotal roles in software validation during containerization before deployment, including code linting (via GitHub Actions), gathering user feedback (through GitHub Issues), and version control. Unit testing can be conducted using GitHub Actions or simulators such as Skymakercam to ensure comprehensive validation and quality assurance of the software.

\subsection{Agile Development} \label{subsec: Agile}
SDSS software development is encouraged to adopt the Agile development methodology, which induces an iterative software development cycle with continuous improvements. Agile development ensures a robust and user-responsive process in which software functionality is segmented into independent testable units. Development commences with the units deemed to be the most critical, and their functionality is verified through unit testing. Subsequently, these units are integrated to perform more complex operations and are tested again. Ultimately, the operations are organized into a single piece of software. Throughout this process, software updates are shared with team members (potential users), and the developer receives feedback from them. This feedback is used to define new units, thereby initiating subsequent development cycles. These cycles recur and evolve until the software quality satisfies the requirements.

\section{LVMAGP Sequence Development} \label{sec:LVMAGP Sequence Development}
\subsection{Autofocus} \label{subsec:Autofocus seq}
\begin{figure} [ht]
   \begin{center}
   \begin{tabular}{c} 
   \includegraphics[width=\textwidth]{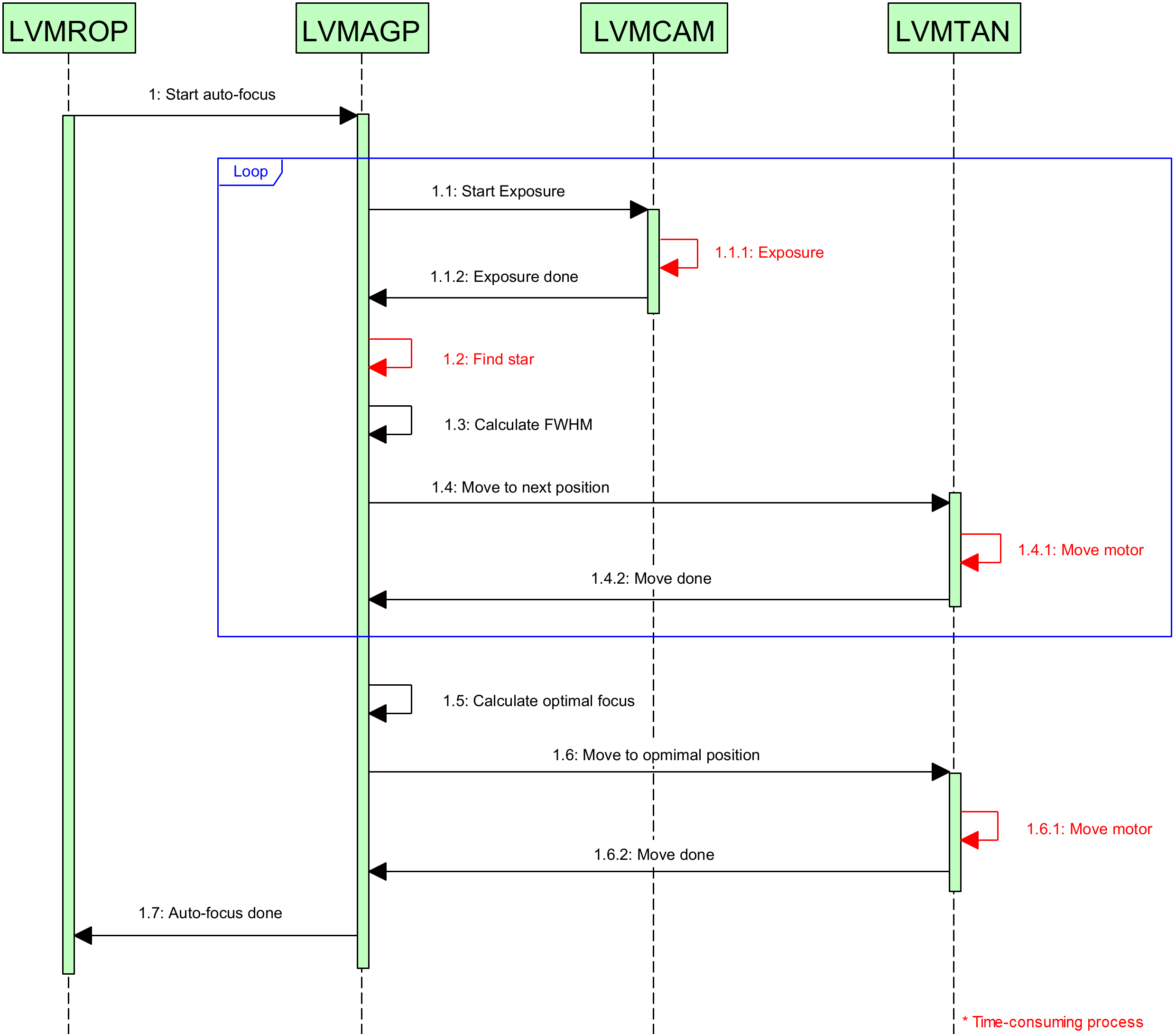}
   \end{tabular}
   \end{center}
   \caption[Sequence diagram for the autofocus sequence] 
   { \label{fig:seqdiag_af} 
Sequence diagram for the autofocus sequence.}
   \end{figure}

The autofocus sequence automatically determines the optimal focus position to implement Req.~8. We assume that the optical fiber and guide camera are in the same focus position; if the guide camera is in focus, then the fiber is as well. This consistency is ensured by the alignment process of the focal plane assembly\cite{Haberle+22}. Autofocus is implemented using the process outlined in Fig.~\ref{fig:seqdiag_af}.

LVMAGP initiates the sequence by taking an exposure to obtain a star image and finds three appropriate stars in the image. Utilizing the DAOFIND algorithm (photutils.detection.DAOStarFinder)\cite{Stetson87,photutils}, the star-finding routine is executed. We filter the detected stars and select the three most appropriate stars in the image. Stars with a high signal-to-noise ratio (SNR) are preferred; however, stars whose maximum pixel value exceeds 60,000 ADU are excluded from the candidate list to avoid saturation. In addition, stars within 20 pixels of the image edge are not considered. One remaining issue is the interruption by neighboring stars. We filter stars that have neighboring stars within five full widths at half maximum (FWHMs) by applying the k-d tree algorithm (scipy.spatial.KDTree)\cite{Maneewongvatana+99, Virtanen+20}.

After selection, each star is fitted using a 2D Gaussian function, resulting in the FWHM in the x and y directions. The average of these FWHMs defines the size of a star, with the representative star size derived from the mean size of the selected trio. The LVMAGP actor records both the focuser encoder position and the corresponding FWHM. By incrementally adjusting the focuser and measuring the star size at each new position, a series of FWHMs is compiled across various focus positions. LVMAGP fits the data in a quadratic function and moves the focuser to the minimum of the fitted function. Quadratic fitting works best near the focus, which is appropriate because the LVMi is a fixed instrument. Temperature change is the main factor for focus change, and it is sufficiently small to be compensated using our autofocus sequence. Hyperbolic fitting can be used to cover a wider focus range.

LVMAGP can also move the focus by following the predefined relationship between temperature and focus position without taking images. The temperature value from the telemetry sensor can be used to compensate for the defocus due to thermal contraction during the night.

\subsection{Field Acquisition} \label{subsec:Field acquisition seq}
\begin{figure} [ht]
    \begin{center}
    \begin{tabular}{c} 
    \includegraphics[width=\textwidth]{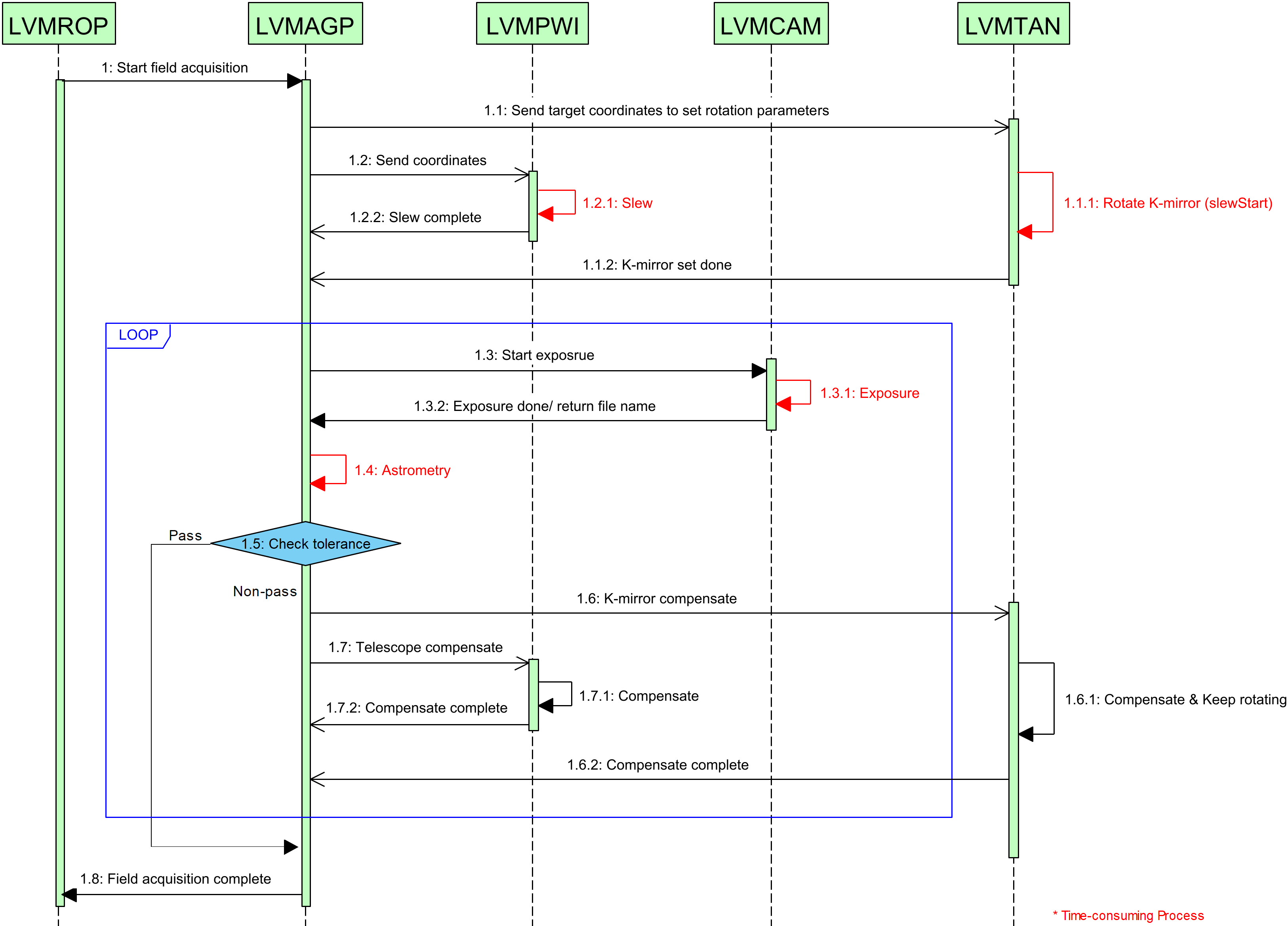}
    \end{tabular}
    \end{center}
    \caption[Sequence diagram for the field acquisition sequence] 
    { \label{fig:seqdiag_fa} 
    Sequence diagram for the field acquisition sequence.}
    \end{figure}

The field acquisition sequence is the process of precisely locating a scientific target at the fiber tip, addressing Reqs. 9--12, as illustrated in Fig.~\ref{fig:seqdiag_fa}. The initial slew is performed based on the internal pointing model (PointXP), whose accuracy is 10 arcseconds in RMS\cite{L350}. The reflection from the siderostat rotates and inverts the coordinates; however, they are corrected by the pointing model. Nevertheless, its accuracy does not meet the Req. 12. Moreover, potential misalignments among the hardware components, such as the siderostat and telescope, can produce slew errors alongside other factors, such as polar alignment.

To compensate for these errors, LVMAGP takes an image immediately after the initial slew and performs astrometry (plate-solving). We installed Astrometry.net\cite{Lang+10} on a local PC with a 4200-index file for plate-solving. Initial guesses regarding the pixel scale, right ascension (R.A.), and declination expedited the astrometry process. Subsequent to astrometry, LVMAGP extracts data from Astrometry.net's output file, including the R.A., declination, and position angle. It then compares the angular distance between the acquired and target fields and compensates for the offset. This process iterates until the offset falls below the defined tolerance (1 arcsecond) to ensure precise field acquisition.

Concurrently, the field rotation is managed by the K-mirror; this functionality is already implemented in the LVMTAN actor. When the telescope starts to slew, LVMTAN calculates the camera field's north direction and its trajectory over time, adjusting the motor speed parameters accordingly. This operation is executed in parallel with the main process by employing a subprocess and asynchronous communication through the CLU.

\subsection{Autoguide} \label{subsec:Autoguide seq}
\begin{figure} [ht]
   \begin{center}
   \begin{tabular}{c} %% tabular useful for creating an array of images 
   \includegraphics[width=\textwidth]{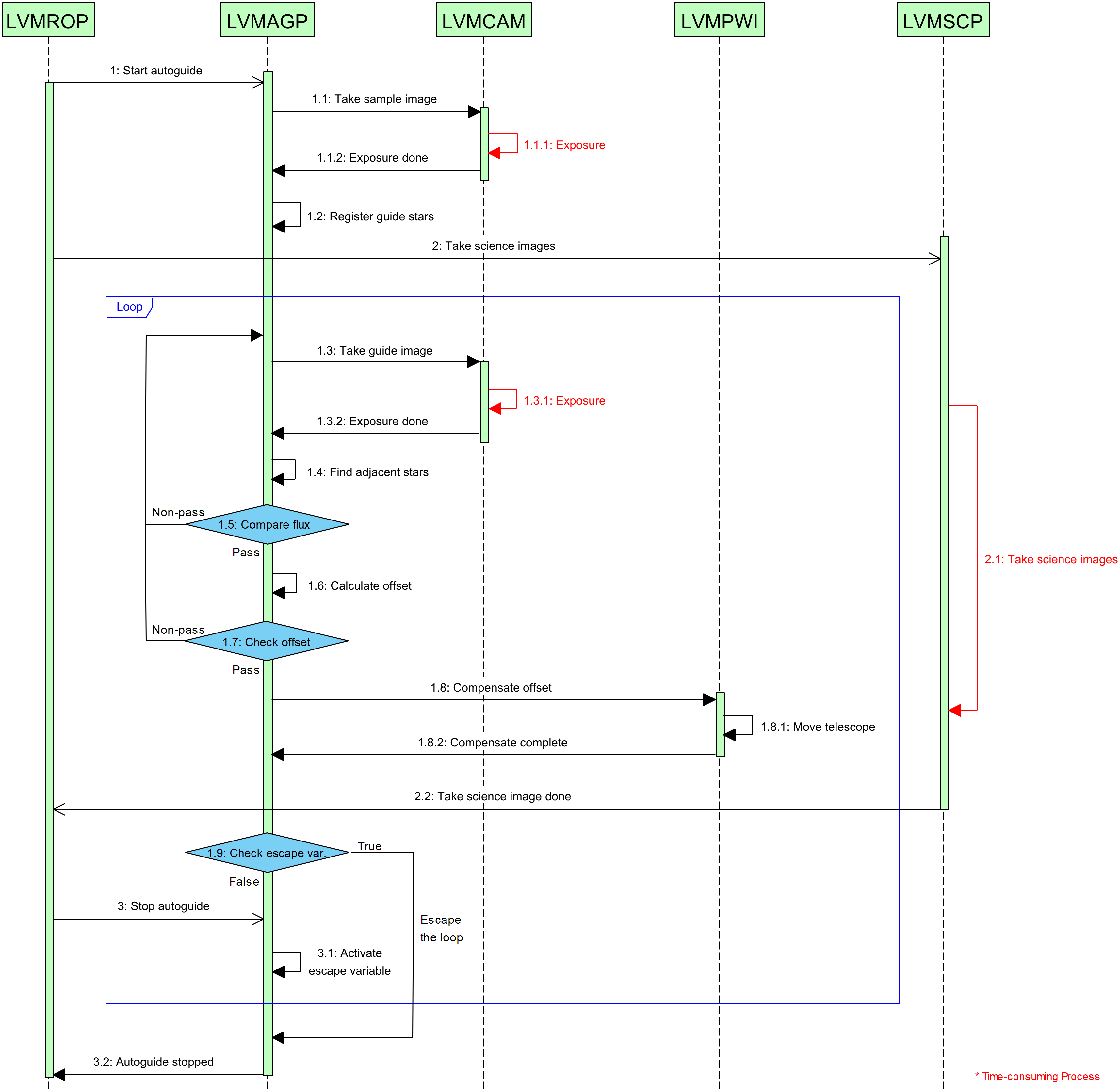}
   \end{tabular}
   \end{center}
   \caption[Sequence diagram for the autoguide sequence] 
   { \label{fig:seqdiag_ag} 
Sequence diagram for the autoguide sequence.}
   \end{figure}

The autoguide sequence maintains the state in which the target is located at the fiber tip. This sequence satisfies Reqs.~10, 12, and 13 through the process illustrated in Fig.~\ref{fig:seqdiag_ag}. LVMAGP selects three isolated bright stars using the same algorithm as the autofocus sequence and registers these stars by recording their locations and fluxes in the image. It then takes another exposure and finds stars in the adjacent 30$\times$30 pixel regions centered around the previously registered stars. Flux is used to identify the newly identified stars as registered stars. A 30\% criterion is then applied to determine whether the adjacent star in the second exposure is the same as that registered in the first exposure. If a discrepancy is observed beyond this criterion, data from the second exposure are discarded, and a new loop begins. Conversely, if they coincide within the tolerance range, LVMAGP calculates the field offset by comparing the star coordinates in the images. The centroid of the triangle formed by the three registered stars becomes the tracking center. LVMAGP evaluates the calculated offset against a criterion of 0.3 pixels to determine the error and the actual field offset. Finally, if the field offset is larger than 0.3 pixels, then LVMAGP sends a command to move the mount according to the results from Eqs.~(\ref{eq:autoguide_dra}) and (\ref{eq:autoguide_ddec}):

    \begin{equation}
    \label{eq:autoguide_dra}
        \Delta\alpha = s \cos\delta (\Delta X \cos\theta + \Delta Y \sin\theta)\\
    \end{equation}
    \begin{equation}
    \label{eq:autoguide_ddec}
        \Delta\delta = s (-\Delta X \sin\theta + \Delta Y \cos\theta) 
    \end{equation}
where $\alpha$ is the R.A., $\delta$ is the declination, $s$ is the pixel scale, $\theta$ is the position angle of the guide image sensor, and $X$ and $Y$ are the coordinates of the guide star in the guide image, respectively.

\section{Proto-model Siderostat} \label{sec:Proto-model Siderostat}
This is the first use of siderostat with the alt-alt configuration. We fabricated a proto-model siderostat to test the LVMAGP software and the feasibility of using the existing pointing model (PointXP) under this unconventional configuration. The pointing error is compensated for by the software; therefore, the requirement for mechanical stability of the proto-model siderostat is relaxed to 6 arcminutes, which is a quarter of the field of view of the guide camera in the LVM.

\subsection{Structure of the Proto-model Siderostat} \label{subsec:Structure of proto-model siderostat}
\begin{figure} [ht]
    \begin{center}
    \begin{tabular}{c} %% tabular useful for creating an array of images 
    \includegraphics[width=\textwidth]{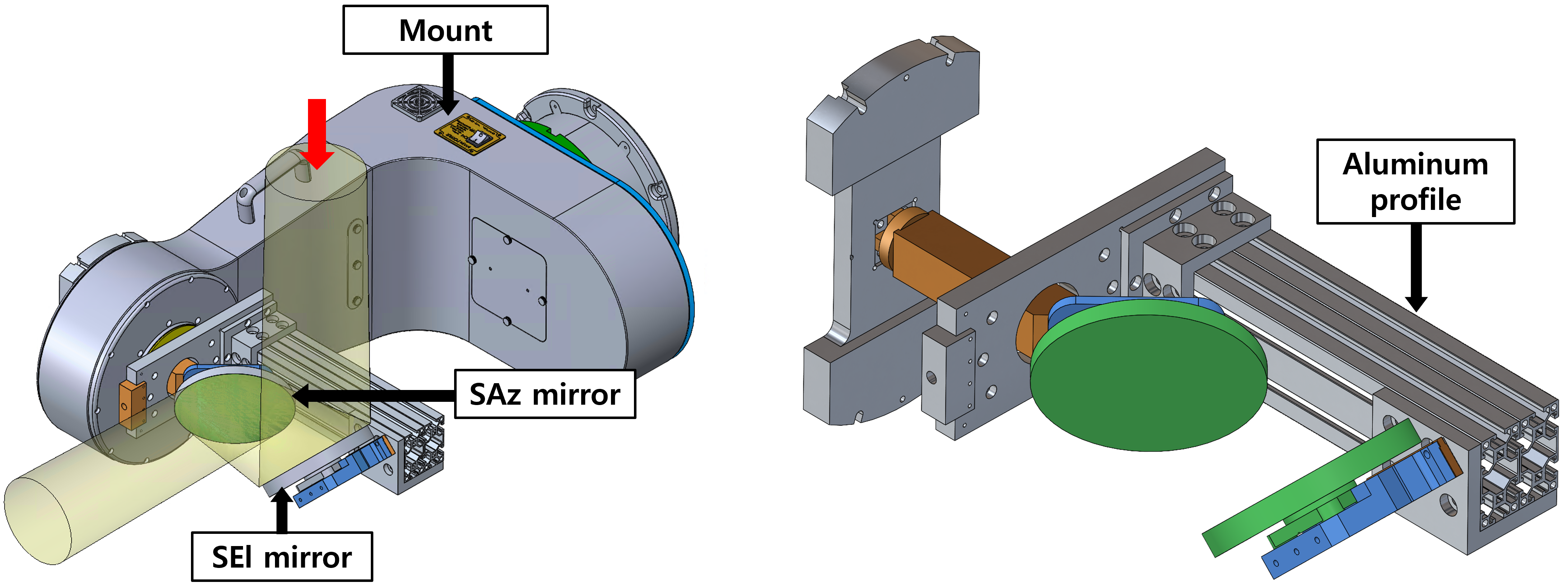}
    \end{tabular}
    \end{center}
    \caption[3D mechanical design of the proto-model siderostat in 3D view] 
    { \label{fig:siderostat_design} 
    3D mechanical design of the proto-model siderostat in 3D view (left) with the mount and (right) without the mount. The SAz and SEl mirrors are colored green and indicated by black arrows. The mirror holders (blue), adapters (orange), and base module (gray) are all assembled. The proto-model siderostat is in the home position, where it points to the zenith. In the left panel, the light path is expressed with light entering direction and indicated with a red arrow. In the right panel, the aluminum profile part is indicated with a black arrow.}
    \end{figure}

The structure should stably support the optical elements within 6 arcmin from gravitational and thermal loads, while also being cost-effective in terms of price, weight, and volume. Figure~\ref{fig:siderostat_design} shows the 3D mechanical design of the proto-model siderostat. The total dimensions are 525 mm (L) $\times$ 292 mm (W) $\times$ 151 mm (H), and the total weight is 9.7 kg. As shown in the left panel of Fig.~\ref{fig:siderostat_design}, the SAz mirror rotates with the azimuth axis of the telescope, whereas the SEl mirror follows the altitude axis. The right panel of Fig.~\ref{fig:siderostat_design} shows the modularized mechanical parts, which have precisely fabricated contact surfaces to achieve high-precision positioning and repeatability by semi-kinematic mounting, resulting in a repeatability within 1 $\mu$m\cite{Slocum10}. Aluminum profiles and commercial couplers are adopted for the SEl mirror base plate and its assembly. The aluminum profiles are made of aluminum alloy 6063-T5, and the other optomechanical parts are fabricated with aluminum alloy 6061-T6 to minimize the difference in the Coefficient of Thermal Expansion (CTE). The SAz and SEl mirror modules are attached to the azimuth and elevation axes of the mount with their own base parts. Two aluminum profiles at 40 mm $\times$ 80 mm $\times$ 350 mm are used for the SEl side base plates.

\subsection{Lightweight and Low-strain Design of the Mirror Module} \label{subsec:Lightweight design of the mirror}
\begin{figure} [ht]
    \begin{center}
    \begin{tabular}{c} %% tabular useful for creating an array of images 
    \includegraphics[width=8cm]{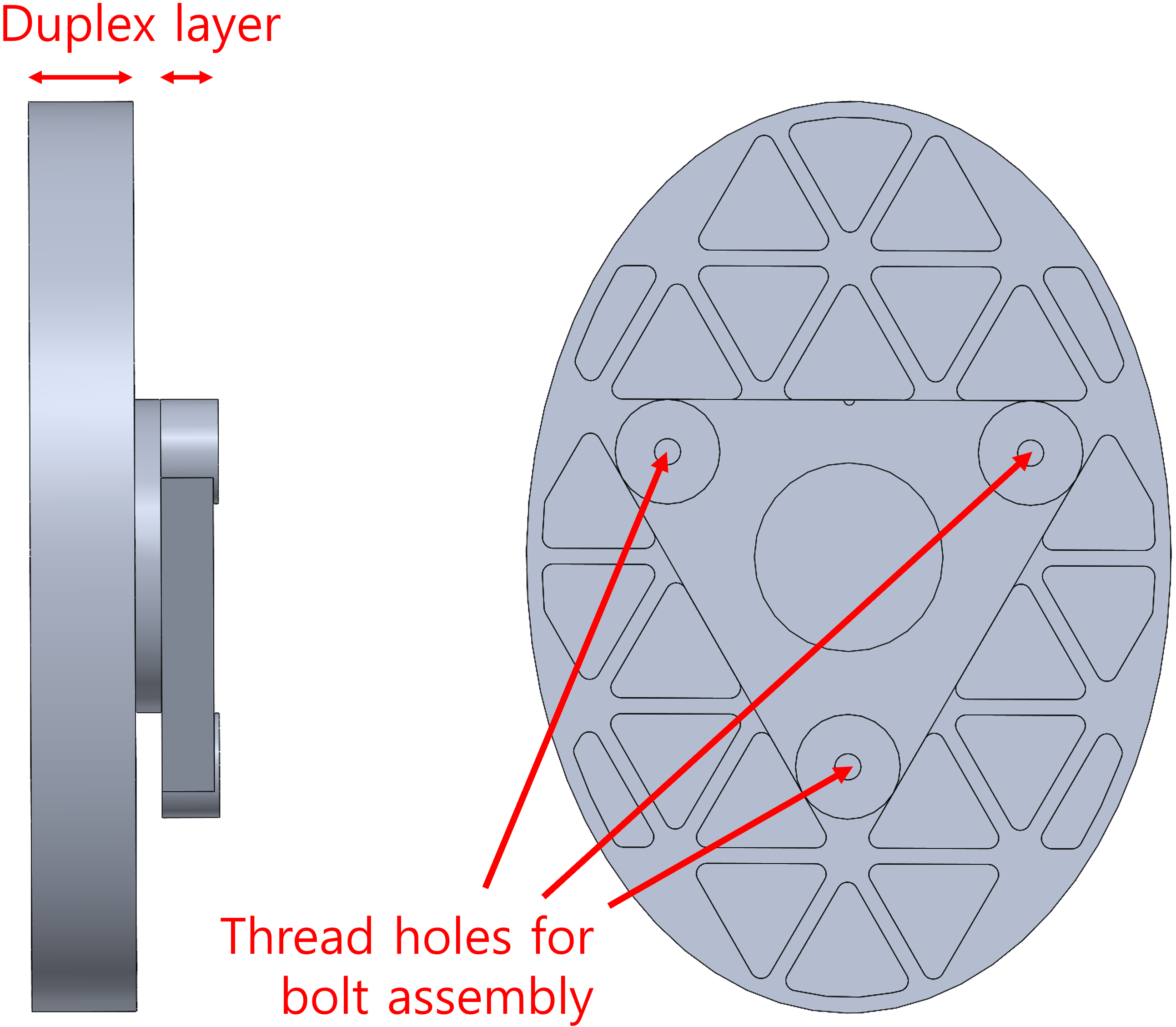}
    \end{tabular}
    \end{center}
    \caption[3D model of the lightweight SAz mirror] 
    { \label{fig:lightweight_mirror} 
    3D model of the lightweight SAz mirror. The left panel shows the side view, and the right panel shows the rear view. The circular neck between the layers separates the mirror into the reflective surface and assembly structure. The assembly structure has three thread holes for the bolt assembly with a mirror holder. The SEl mirror shares the same design concept and structure as the SAz mirror.}
    \end{figure}

The siderostat includes two flat mirrors: the SEl mirror, with a circular aperture and diameter of 174 mm, and the SAz mirror, with an elliptical aperture and major and minor axes of 174 mm and 123 mm, respectively. Both mirrors are made of the aluminum alloy 6061-T6, which is the same material used for optomechanics. A duplex-layer design of the mirrors separates the reflecting surface from the assembly structure; thus, the optical surface deformation caused by the mounting stress is mitigated (Fig.~\ref{fig:lightweight_mirror}). The assembly structure of the mirror has a triangular shape that fits in the grooves of the mirror holder\cite{Han+21}. The fabrication accuracy of the grooves is 50 $\mu$m, and the mirrors can be placed within a precision of $\pm$1 $\mu$m\cite{Slocum10}.

The mirror is lightweight with isogrid pockets in an open-back style that is less solid relative to a solid mirror; however, it takes a short time to reach thermal equilibrium compared to the solid and sandwich structure\cite{Park+22TRT, Ahmad+18}. The mass of the lightweight mirror is 643 g for the SAz mirror and 774 g for the SEl mirror, with a lightweight rate of 46\%. The designed mirrors were machined using a Single-Point Diamond Turning Machine (SPDTM).

\subsection{Deflection and Stress Simulations} \label{subsec:Deflection and Stress Simulations}
Siderostat optomechanics is deflected by self-weight, bolting torque, and thermal load, which increases the scattering of the pointing model. To evaluate the stability of the siderostat, we performed a static analysis using the Finite Element Method (FEM) in SolidWorks (Dassault Syst\`{e}ms)\cite{Solidworks} at four positions: zenith (home), east horizon, and two north horizons. The north horizon is a singular point of the north-south alt-alt mount, similar to the zenith for the alt-az mount. As long as one axis heads toward the north horizon, the other axis can be rotated at any angle. Two angles of the free axis for the east horizon and meridian were selected.

For the Finite Element Analysis (FEA), gravitational acceleration at 1G is applied in the --Z-axis direction, and the reference temperature at zero strain is set to 25$^\circ$C, which is the annual maximum temperature at LCO\cite{Thomas+10}. A thermal load of $\Delta T=-30^\circ$C (corresponding to $T=-5^\circ$C) was applied and compared with the case of zero thermal load to simulate the annual temperature range at LCO. In the simulation, the contact surfaces between the same materials are considered bonded, while the interfaces between different materials are considered to be in contact. The bolting torque is 2 Nm, and a 3D hyperbolic tetrahedral solid mesh with a size range of 7 to 130 mm is applied. The numbers of nodes and mesh elements vary depending on the position, and the medians are 892,506 and 480,517, respectively.

\begin{figure} [ht]
    \begin{center}
    \begin{tabular}{c} 
    \includegraphics[width=\textwidth]{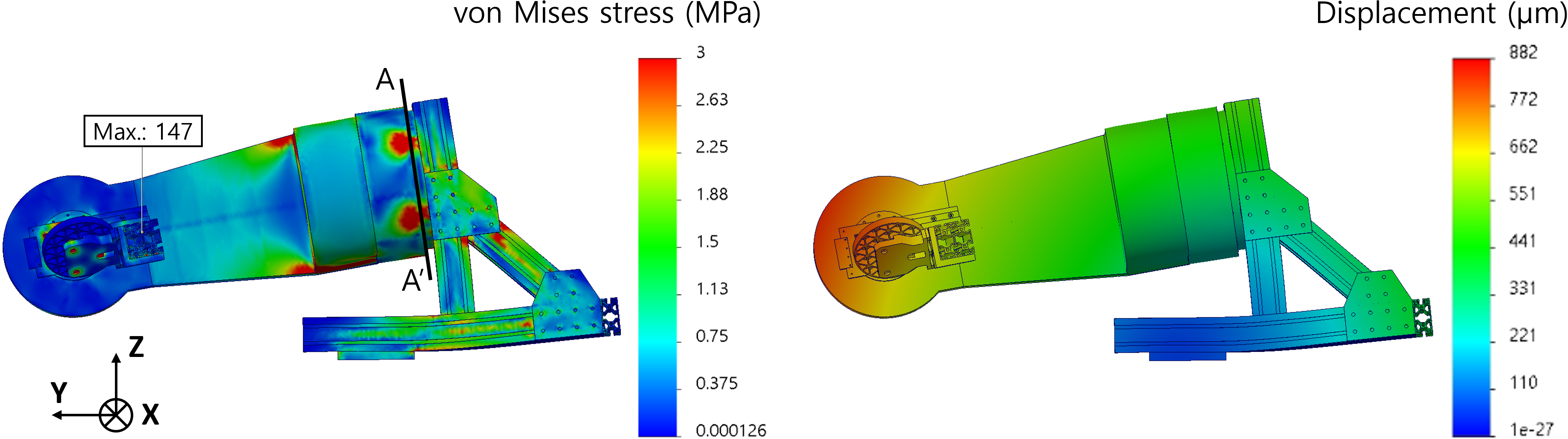}
    \end{tabular}
    \end{center}
    \caption[Stress and displacement of the proto-model siderostat at the zenith position at the temperature of $-5^\circ$C]
    { \label{fig:FEA_optomech} 
    (left) Loaded von Mises stress and (right) displacement of the proto-model siderostat at the zenith position at the temperature of $-5^\circ$C. The figures are 300 times more exaggerated than the actual variations. The maximum stress point is indicated in the left panel; however, this point is a stress singularity and should be excluded from the evaluation of FoS. Actual maximum stress is applied by bolting torque on section A-A$'$, where the wedge and mount are connected by bolting.}
    \end{figure}

Figure~\ref{fig:FEA_optomech} shows the FEA results for the siderostat in the zenith position at $T=-5^\circ$C. The maximum von Mises stress is 93.3 MPa when the siderostat points to the zenith, and it corresponds to a factor of safety (FoS) of 2.3, which is calculated using Eq.~(\ref{eq:FoS}). The rule of thumb gives a target FoS based on the uncertainty of the material properties, load stress, geometry, failure analysis, and reliability; in our case, the target FoS is 1.7\cite{Ullman17}. The weight of the mount bends the wedge, yielding the $\alpha$-tilt and Z- and Y-decenters (Tables~\ref{tab:FEA_grav} and \ref{tab:FEA_thermal}). The mass imbalance along the X-axis causes a little X-decenter and $\beta$- and $\gamma$-tilts.
    \begin{equation}
    \label{eq:FoS}
        \text{FoS} = \frac{\text{Allowable strength}}{\text{Applied stress}} = \frac{\text{Yield strength}}{\text{Maximum von Mises stress}}
    \end{equation}

Tables~\ref{tab:FEA_grav} and \ref{tab:FEA_thermal} also present the decenters and tilts at different positions and temperatures. In all cases, the main deflection is from the Z-decenter and $\alpha$-tilt, and the tendency of deflection resembles that in Fig.~\ref{fig:FEA_optomech}. To evaluate the effect on the pointing accuracy, we only compare the $\alpha$-tilt because the displacement of flat mirrors does not affect the accuracy of the pointing and the other tilts are much smaller than the $\alpha$-tilt. Then, according to the largest difference in the $\alpha$-tilt, the maximum scatter of the pointing model induced by the tilt of optomechanics is expected to be 4.4 arcminutes for the extreme case, which can be compensated for by using our field acquisition sequence.

\begin{figure} [ht]
    \begin{center}
    \begin{tabular}{c} 
    \includegraphics[width=\textwidth]{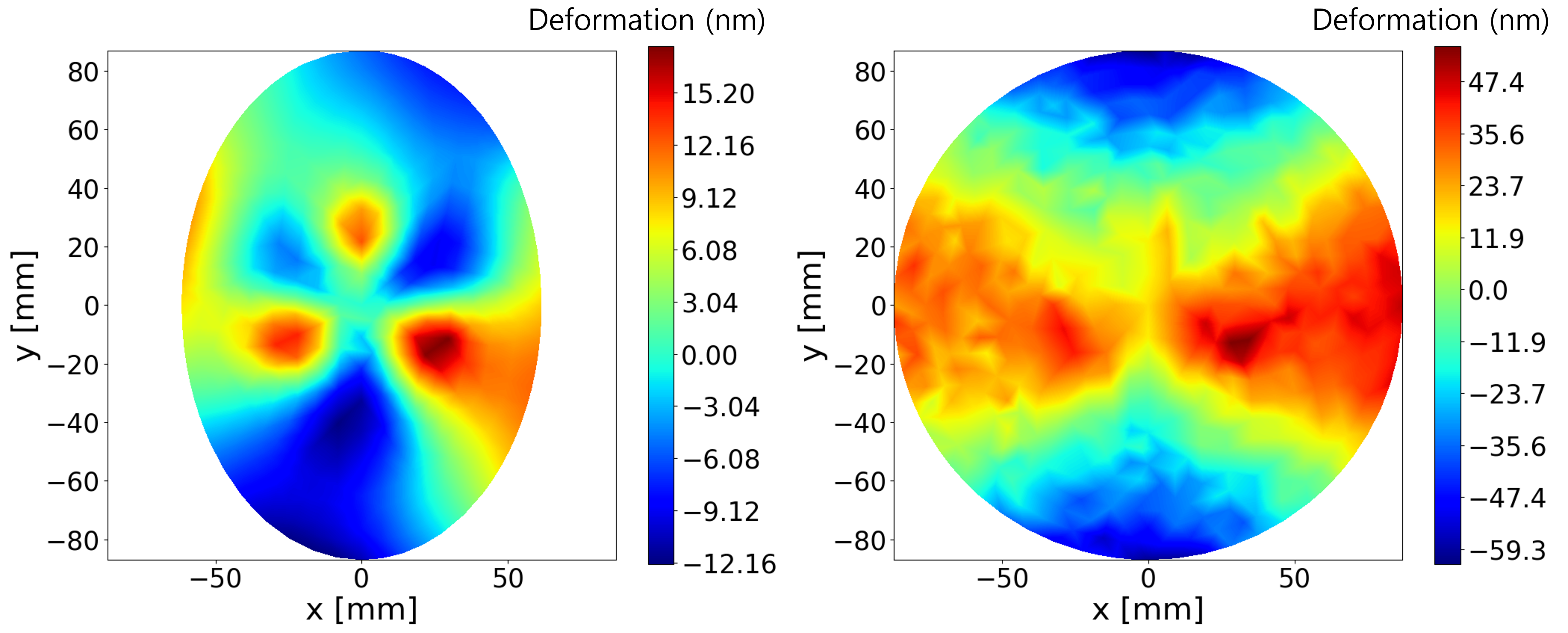}
    \end{tabular}
    \end{center}
    \caption[Deformation of the reflective surfaces at the zenith position at the temperature of $-5^\circ$C]
    { \label{fig:FEA_mirror} 
    Deformation of the reflective surfaces at the zenith position at the temperature of $-5^\circ$C: (left) SAz mirror and (right) SEl mirror. The noisy high-frequency pattern is due to a relatively large mesh size and not a real deformation.}
    \end{figure}

Figure~\ref{fig:FEA_mirror} illustrates the deformation of the mirror surface attributed to the combined effects of self-weight, bolting torque, and thermal load at the zenith position. The SEl mirror shows a larger deformation than the SAz mirror because of its large size and heavy weight. For the SAz mirror, the three points corresponding to the bolting positions are deformed by 16 nm, and an astigmatism pattern is also observed. Moreover, for the SEl mirror, a gravitational effect arises, and defocus is observable with the astigmatism. Tables~\ref{tab:FEA_grav} and \ref{tab:FEA_thermal} list the surface deformation for all simulated cases in P--V and RMS. The deformations calculated from these simulations are smaller than the surface figure errors presented in Sec. \ref{subsec:Lightweight design of the mirror} by one order of magnitude.

    \begin{table}[ht]
    \caption{FEA results of the proto-model siderostat ($T=25^\circ$C)}
    \label{tab:FEA_grav}
    \begin{center}
    \begin{tabular}{cc|ccc|ccc|cc|c}
    \hline
    \rule[-1ex]{0pt}{3.5ex}  Position & Mirror & \multicolumn{3}{|c|}{Decenter} & \multicolumn{3}{|c|}{Tilt$^{a}$} & \multicolumn{2}{|c|}{Surface deformation} & FoS\\
    \rule[-1ex]{0pt}{3.5ex}   & & \multicolumn{3}{|c|}{($\mu$m)} & \multicolumn{3}{|c|}{(arcsecond)} & \multicolumn{2}{|c|}{(nm)} & \\
    \cline{3-10}
    \rule[-1ex]{0pt}{3.5ex}   & & X & Y & Z & $\alpha$ & $\beta$ & $\gamma$ & ~~P--V$^{b}$~~ & RMS & \\
    \hline
    \rule[-1ex]{0pt}{3.5ex}  Zenith & SAz & 39 & 113 & -412 & -102.8 & 4.8 & 5.2 & 27.0 & 5.7 & 10.8\\
    \rule[-1ex]{0pt}{3.5ex}         & SEl & 30 & 104 & -401 & -112.0 & 8.4 & 15.3 & 55.5 & 13.7 & \\
    \hline
    \rule[-1ex]{0pt}{3.5ex}  East & SAz & 1 & 76 & -539 & -162.7 & 0.6 & 0.3 & 21.6 & 4.2 & 8.4\\
    \rule[-1ex]{0pt}{3.5ex}       & SEl & 1 & -58 & -545 & -172.8 & 1.3 & -3.6 & 53.9 & 12.6 &\\
    \hline
    \rule[-1ex]{0pt}{3.5ex}  North 1$^{c}$ & SAz & 40 & 119 & -432 & -108.0 & 5.3 & 4.0 & 8.6 & 1.9 & 7.8\\
    \rule[-1ex]{0pt}{3.5ex}                & SEl & 25 & 122 & -439 & -111.4 & -12.1 & 0.4 & 52.9 & 12.2 & \\
    \hline
    \rule[-1ex]{0pt}{3.5ex}  North 2$^{c}$ & SAz & 1 & 81 & -490 & -149.0 & 2.6 & 0.7 & 21.6 & 4.9 & 9.8\\
    \rule[-1ex]{0pt}{3.5ex}                & SEl & 12 & -34 & -498 & -146.8 & -12.6 & -1.7 & 31.6 & 8.0 & \\
    \hline
    \end{tabular}
    \end{center}
    {$^{a}$ $\alpha$-, $\beta$-, and $\gamma$-tilts are angles of rotation with X-, Y-, and Z-axes defined in Fig.~\ref{fig:FEA_optomech}, respectively.\\
    $^{b}$ Peak-to-valley\\
    $^{c}$ North 1 and North 2 are the positions where the azimuth coordinate line of the mount passes through the zenith and east positions, respectively.}
    \end{table}
    
    \begin{table}[ht]
    \caption{FEA results of the proto-model siderostat ($T=-5^\circ$C)}
    \label{tab:FEA_thermal}
    \begin{center}
    \begin{tabular}{cc|ccc|ccc|cc|c}
    \hline
    \rule[-1ex]{0pt}{3.5ex}  Position & Mirror & \multicolumn{3}{|c|}{Decenter} & \multicolumn{3}{|c|}{Tilt} & \multicolumn{2}{|c|}{Surface deformation} & FoS\\
    \rule[-1ex]{0pt}{3.5ex}   & & \multicolumn{3}{|c|}{($\mu$m)} & \multicolumn{3}{|c|}{(arcsecond)} & \multicolumn{2}{|c|}{(nm)} & \\
    \cline{3-10}
    \rule[-1ex]{0pt}{3.5ex}   & & X & Y & Z & $\alpha$ & $\beta$ & $\gamma$ & ~~~P--V~~~ & RMS & \\
    \hline
    \rule[-1ex]{0pt}{3.5ex}  Zenith & SAz & 90 & -191 & -716 & -103.2 & 6.1 & 7.7 & 28.9 & 5.9 & 2.3\\
    \rule[-1ex]{0pt}{3.5ex}         & SEl & 166 & -223 & -702 & -113.5 & 11.0 & 30.0 & 105.0 & 25.2 & \\
    \hline
    \rule[-1ex]{0pt}{3.5ex}  East & SAz & 1 & -233 & -780 & -162.8 & 0.2 & 0.2 & 33.2 & 6.9 & 2.2\\
    \rule[-1ex]{0pt}{3.5ex}         & SEl & 1 & -385 & -702 & -185.2 & 0.6 & -4.7 & 77.1 & 18.7 &\\
    \hline
    \rule[-1ex]{0pt}{3.5ex}  North 1 & SAz & 89 & -189 & -725 & -105.6 & 6.1 & 7.9 & 28.9 & 5.9 & 2.3\\
    \rule[-1ex]{0pt}{3.5ex}          & SEl & 157 & -185 & -760 & -109.8 & -32.6 & 1.7 & 130.8 & 30.0 & \\
    \hline
    \rule[-1ex]{0pt}{3.5ex}  North 2 & SAz & 3 & -237 & -778 & -165.4 & 2.2 & 0.1 & 34.7 & 6.6 & 2.4\\
    \rule[-1ex]{0pt}{3.5ex}          & SEl & 31 & -360 & -702 & -161.2 & -26.2 & -4.8 & 44.1 & 10.9 & \\
    \hline
    \end{tabular}
    \end{center}
    \end{table}

\section{Performance Test} \label{sec:performance}
\subsection{Instrument Setup} \label{subsec:Instrument setup}
\begin{figure} [ht]
   \begin{center}
   \begin{tabular}{c} %% tabular useful for creating an array of images 
   \includegraphics[width=0.9\textwidth]{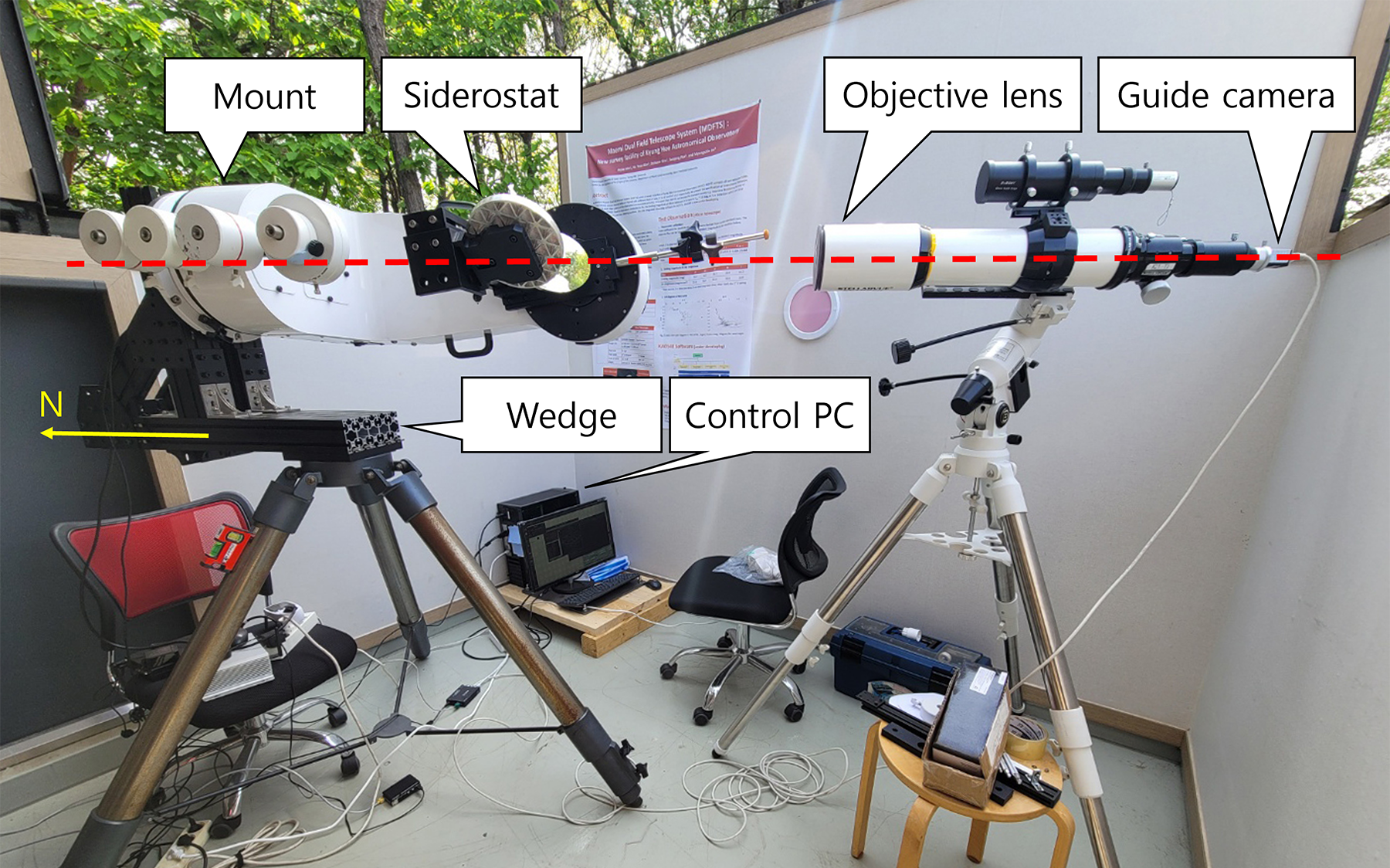}
   \end{tabular}
   \end{center}
   \caption[Performance test setup with the fabricated siderostat] 
   { \label{fig:testsetup} 
Performance test setup with the fabricated siderostat. The mount, proto-model siderostat, wedge, control PC, objective lens, and guide camera are labeled. The north direction is indicated by a yellow arrow and a yellow ``N" character. The optical axis of the objective lens is a light-entering direction towards the objective lens, and it is indicated by the red dashed line. Same as in Fig.~\ref{fig:siderostat_design}, the proto-model siderostat is in the home position, where the telescope points to the zenith.}
   \end{figure}

The proto-model siderostat was installed at the Kyung Hee Astronomical Observatory (KHAO) in Yongin, South Korea. The geographical coordinates of the installation site are as specified 37:14:20.7N, 127:04:56.1E, and 119 m altitude\cite{Im+21}. Figure~\ref{fig:testsetup} shows the experimental setup for the LVMAGP performance test, which consists of the siderostat and a telescope with an image sensor. During the installation process, the wedge is leveled and aligned in the north-south direction. The mount stands vertical to the wedge to operate the alt-azimuth mount as an alt-alt mount. Counterweights were attached to both axes to maintain balance. For imaging optics, a Stellarvue 102 mm apochromatic refractor (f/7) is used, whose optical axis is aligned with the azimuth axis of the mount. The focuser and the K-mirror were not included in the test setup. In their place, mock devices simulated by the LVMTAN and Skymakercam actors are used. Consequently, the guide camera, which is identical in model to that in the LVMi, is directly mounted onto the telescope without the focal plane assembly fixed to the optical table. Data cables from both the mount and guide camera are connected to the control PC. The overall hardware and software configuration is the same as that in Fig.~\ref{fig:architecture_allinst}, although it only includes elements related to the ``sci" telescope.

\subsection{Test Results} \label{subsec:Test results}
\subsubsection{Autofocus} \label{subsubsec:Autofocus result}
As mentioned in Sec.~\ref{subsec:Instrument setup}, conducting a real hardware test for the autofocus sequence was not possible because of the absence of a motorized focuser. To overcome this limitation, we used the Skymakercam and the LVMTAN simulator as alternatives to evaluate the autofocus sequence. As outlined in Sec.~\ref{subsec:LVMi System Architecture}, Skymakercam can generate realistic images that more accurately replicate the actual images at the LCO than those obtained by manually adjusting the position of the guide camera through the focuser.

\begin{figure} [ht]
   \begin{center}
   \begin{tabular}{c} %% tabular useful for creating an array of images 
   \includegraphics[width=\textwidth]{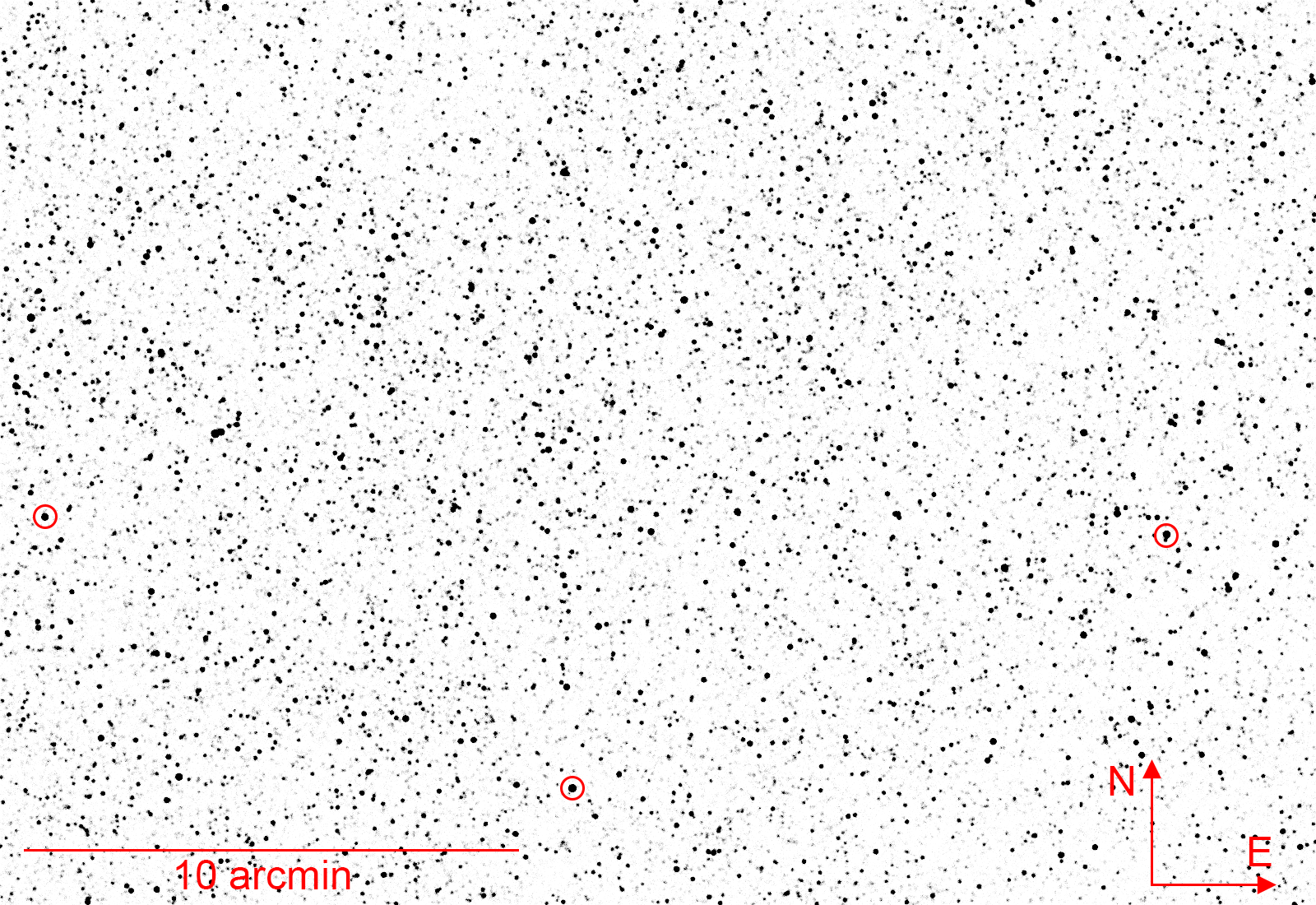}
   \end{tabular}
   \end{center}
   \caption[Synthetic guide camera image of example crowded field]
   { \label{fig:starfind} 
Synthetic guide camera image of example crowded field ($\alpha=18^{\text{h}}\ 00^{\text{m}}\ 15^{\text{s}},\ \delta=-29^\circ\ 03'\ 02''$). The image is inverted after square-root-scaling and shown with compass arrows and scale bar. Three stars selected by star finding algorithm are marked with a red circle. Although another star is located 2.2 FWHM apart from the rightmost marked star, it does not affect the performance because it is 80 times dimmer than the bright star.}
   \end{figure}

\begin{figure} [ht]
   \begin{center}
   \begin{tabular}{c} %% tabular useful for creating an array of images 
   \includegraphics[height=7cm]{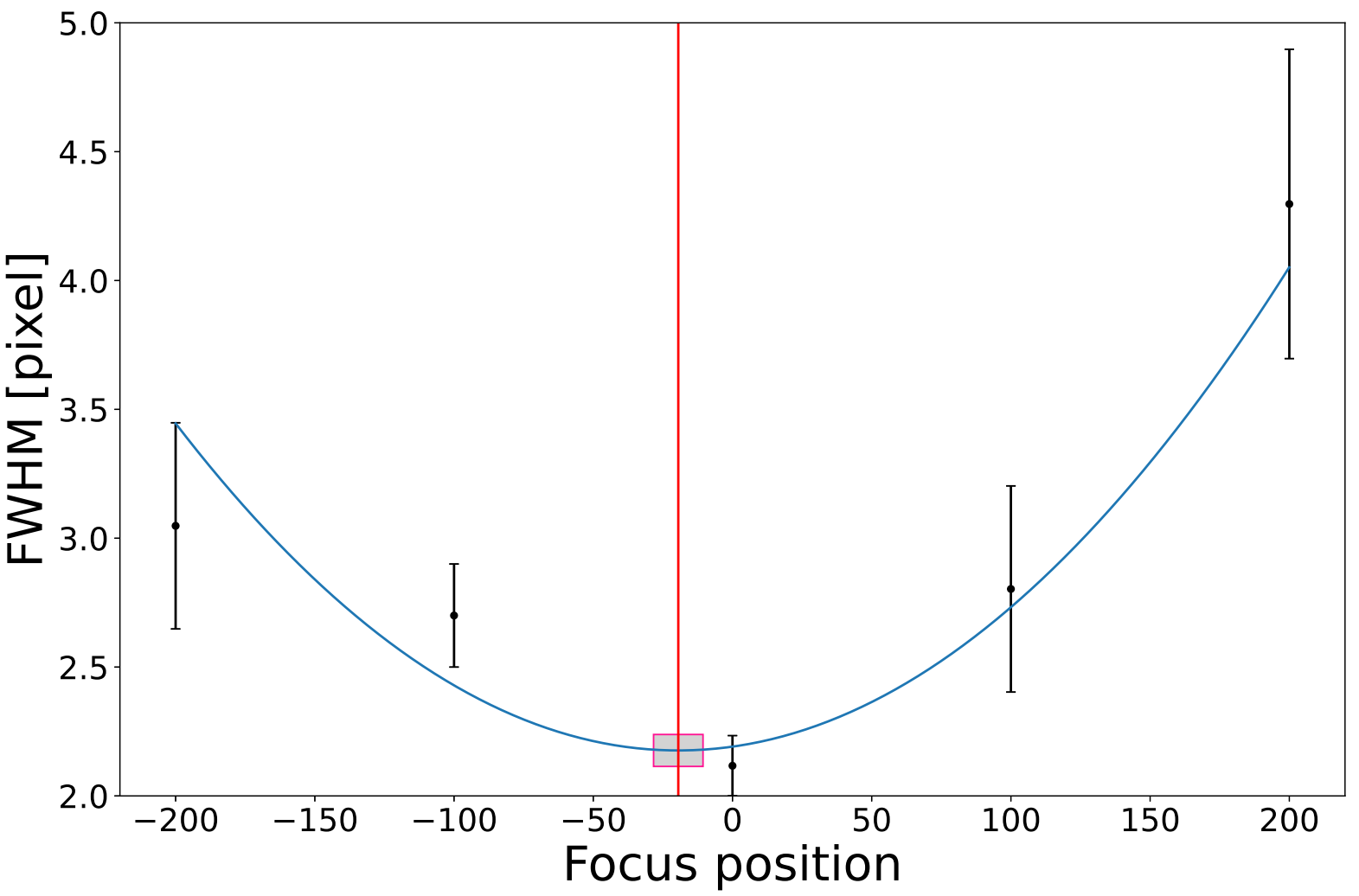}
   \end{tabular}
   \end{center}
   \caption[FWHM of stars in terms of the focus position] 
   { \label{fig:focus_FWHM} 
FWHM of stars in terms of the focus position. The error bars are the sample standard deviations of FWHM of the measured stars at each focus position. The blue curve is the fitted line, and the red vertical line means the optimal focus position. The pink box represents the optical focus reflecting the error range.}
   \end{figure}

Figure~\ref{fig:starfind} displays a synthetic guide image generated by Skymakercam for a crowded field. The three selected stars are the results of the star-finding algorithm, and they satisfy the criteria described in Sec.~\ref{subsec:Autofocus seq}. Figure~\ref{fig:focus_FWHM} shows the results of the autofocus sequence. LVMAGP measures the FWHM of stars at various focus positions and determines the optimal focus using quadratic polynomial fitting. Consequently, the autofocus sequence enables LVMAGP to successfully satisfy Req.~8.

\subsubsection{Field acquisition} \label{subsubsec:Field acquisition result}
Prior to evaluating the field acquisition sequence, we established the pointing model. Using 30 stars, the pointing model converged with an RMS error of 9 arcminutes, showing the applicability of this pointing model to the alt-alt configuration employed by the siderostat. However, this is a much larger value than the expected accuracy of the pointing model (10 arcseconds). This discrepancy was primarily due to the misalignment between the optical axis of the telescope and the rotation axes of the proto-model siderostat mirror and mount. This exercise revealed a limitation: the pointing model cannot compensate for such an error because it does not occur in standard alt-azimuth or equatorial configurations where the telescope is directly mounted, and the model is not designed for this unconventional application. Nonetheless, this error was covered by astrometry in the sequence, thus satisfying Req.~11.

\begin{figure} [ht]
   \begin{center}
   \begin{tabular}{c} %% tabular useful for creating an array of images 
   \includegraphics[width=\textwidth]{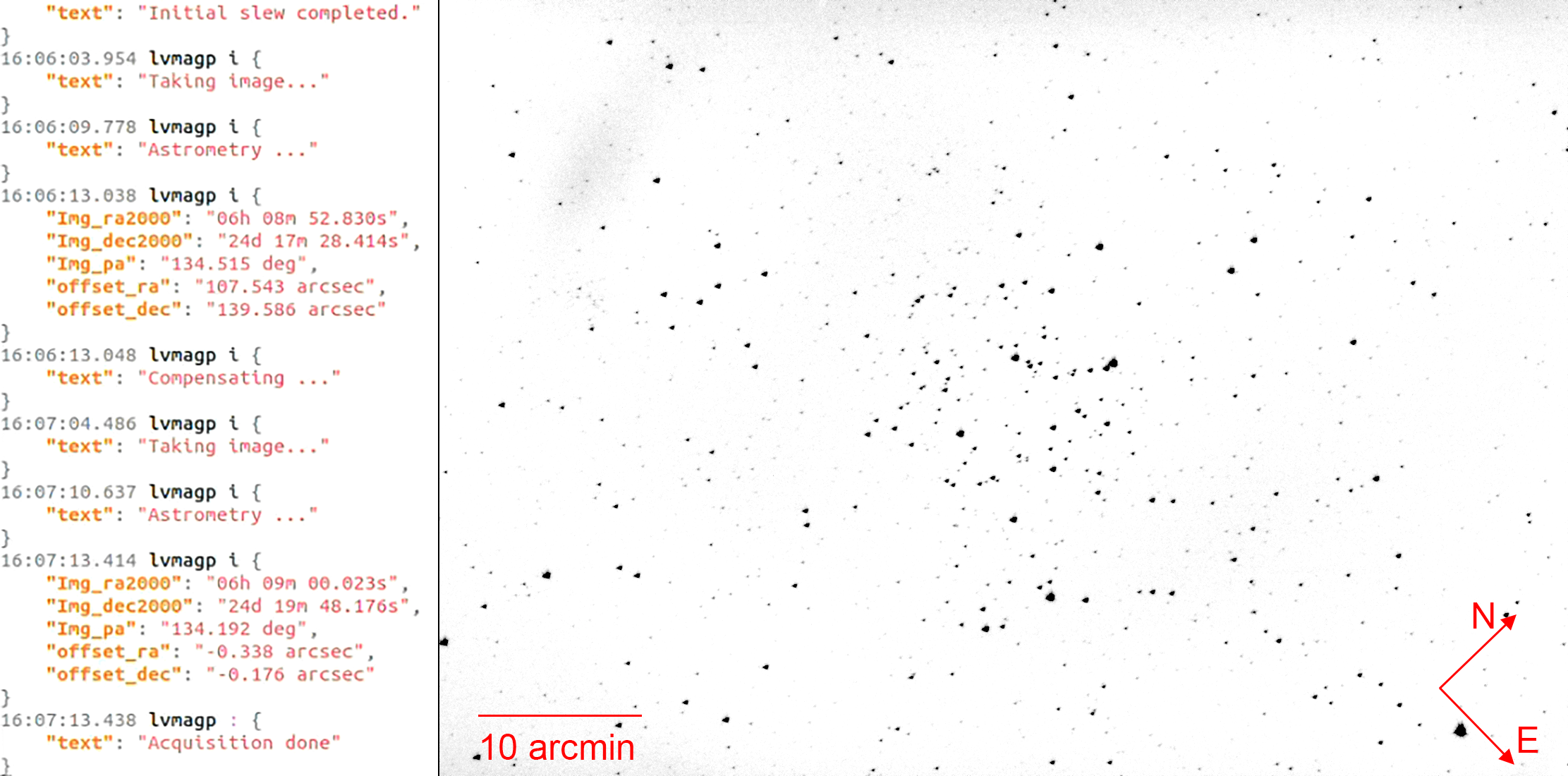}
   \end{tabular}
   \end{center}
   \caption[Implementation of the field acquisition sequence] 
   { \label{fig:field_acquisition} 
(left) Actor prints during the field acquisition command. (right) Acquired image of Messier 35 after the command in the left panel. The image is shown with compass arrows and scale bar and is not flat-fielded.}
   \end{figure}

Figure~\ref{fig:field_acquisition} shows the performance of the field acquisition sequence by acquiring an open cluster, Messier 35, as a target. The initial slew errors in the equatorial coordinates were substantial ($\Delta\alpha=-107''.543$ and $\Delta\delta=139''.58$). However, after a single round of compensation, the final acquisition was achieved with a markedly reduced error of $\Delta\alpha=-0''.33$ and $\Delta\delta=-0''.18$, for a total error of $0''.38$. Messier 35 was acquired successfully, as shown in the right panel of Fig.~\ref{fig:field_acquisition}. Despite the large initial slew error, the field acquisition sequence worked flawlessly, and LVMAGP satisfied Reqs.~9--11 and half of Req.~12. The compensation process required 70 s for completion. Most of this time was spent waiting for the mount to stabilize after the slew. This duration is expected to decrease significantly during actual commissioning with a real siderostat through careful balancing of the mount's axes.

\subsubsection{Autoguide} \label{subsubsec:Autoguide result}
For the autoguide sequence test, owing to field rotation, we could maintain the autoguide for only 7 min, which is almost half the regular exposure of the LVM survey (15 min). Throughout this period, 101 compensations were performed with an autoguiding exposure of 3 s. This number of compensations was assumed to be sufficient, and extending the duration of the observation was assumed to have an insignificant impact on the guide error statistics.

\begin{figure} [ht]
   \begin{center}
   \begin{tabular}{c} %% tabular useful for creating an array of images 
   \includegraphics[width=\textwidth]{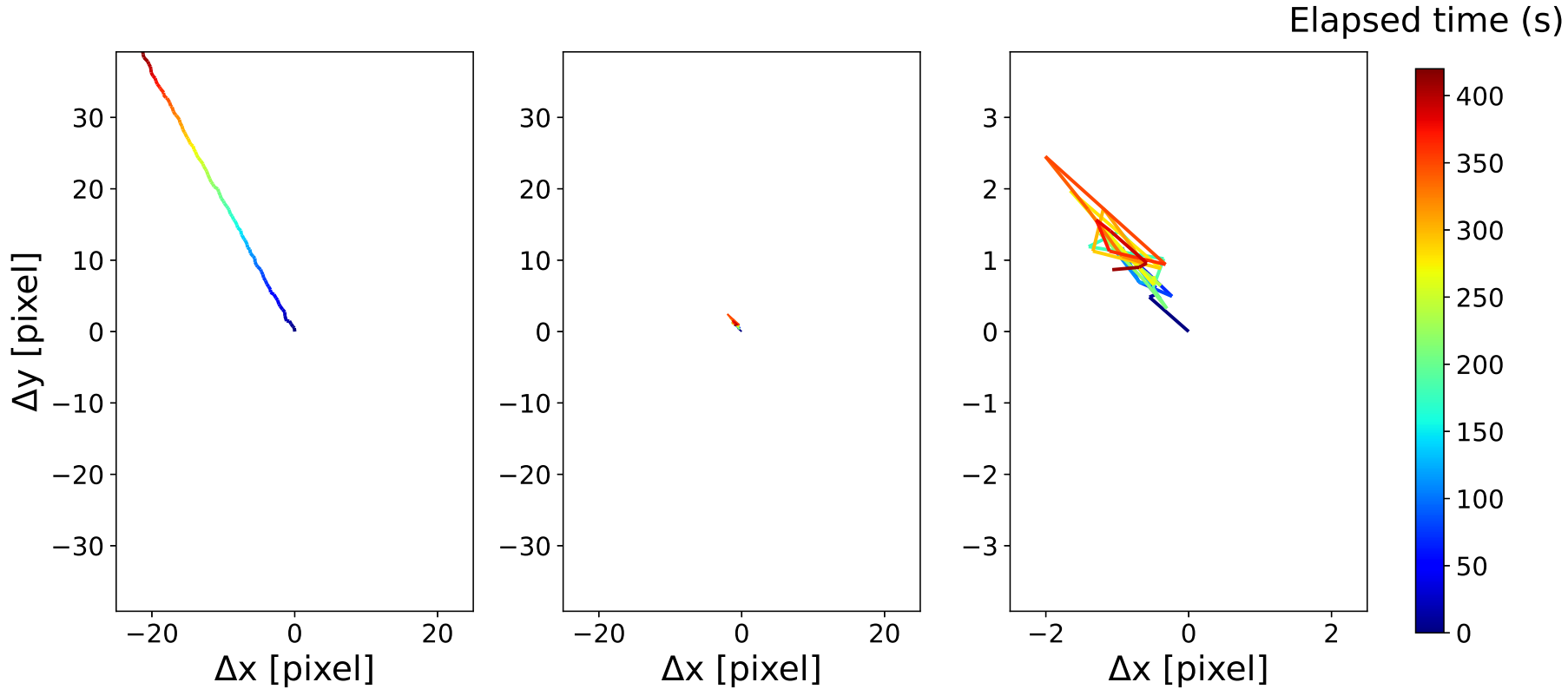}
   \end{tabular}
   \end{center}
   \caption[Movement of the tracking center as time flows] 
   { \label{fig:autoguide_comp}
Movement of the tracking center as time flows (left) without autoguiding and (middle) with autoguiding. The right panel is a 10x zoomed graph of the middle panel. The color bar applies to all panels.}
   \end{figure}

The field rotation occurred since we performed the test without the K-mirror. To separate the field rotation and tracking error, we define the tracking center as the centroid of a triangle formed by the three guide stars. If there is no tracking error, the field should rotate around the fixed tracking center. We track its movement over time when the autoguiding is on and off. Figure~\ref{fig:autoguide_comp} indicates that the drift of the tracking center decreases dramatically with the implementation of autoguiding. Without autoguiding, the tracking center drifted by 13 pixels (34 arcseconds) in RMS. This drift was reduced to 0.59 pixels (1.5 arcseconds) in RMS with autoguiding. However, the improved drift still exceeded the limit of Req. 12 ($<$1 arcsecond).

In the right panel of Fig.~\ref{fig:autoguide_comp}, the switchback motion restores the shift, but the tracking center moves slowly in one direction, which is also observed in the non-autoguide case. This is because the tracking error produced by the error of the pointing model is nonnegligible inside one autoguiding loop. In contrast to Sec. \ref{subsubsec:Field acquisition result}, the error of the pointing model affects the performance. This problem should be resolved by more precise alignment among the hardware elements, making a better pointing model. Notably, when the LVMi is installed at the LCO, we expect the point model to be improved based on a more precise alignment using a pier and an Alignment Autocollimator\cite{Herbst+20}. Blind pointing and tracking performance are expected to be at least 10 times higher than those obtained in our test results, thereby solving the fundamental problem observed during testing.
   
\section{Conclusion} \label{sec:conclusion}
The LVM project requires the continuous control of four telescope units. To operate the LVMi telescope subsystem, we developed the telescope control software LVMAGP, which has a hierarchical architecture and three key functionalities (autofocus, field acquisition, and autoguide) following the SDSS framework. The adoption of an asynchronous CLU and containerization enabled the parallel operation of multiple instruments and telescope units. The Agile development methodology was used to efficiently develop and improve the software within the evolving release cycle. Moreover, we implemented three critical functions within the \ttfamily LVMTelescopeUnit\rmfamily\; class by defining the sequences. Since LVMAGP provides an API for higher-level software, LVMROP will be able to configure higher-level functions for robotic observation using our API.

Before integrating the LVMi, a proto-model siderostat was designed and fabricated with duplex-layered lightweight mirrors and precise contact points. FEA of the deflection due to self-weight, bolting torque, and thermal deformation predicted that mechanical errors can be compensated for by the LVMAGP. We evaluated whether LVMAGP fulfilled the requirements using simulators and a proto-model siderostat. Performance tests revealed that LVMAGP met all specified requirements except for Req.~12. The autoguiding RMS error exceeded 1 arcsecond owing to a misalignment between the telescope and mount. This problem is expected to be resolved by more precise alignment among the hardware components at the LCO.

\subsection* {Acknowledgments}
Funding for the Sloan Digital Sky Survey V has been provided by the Alfred P. Sloan Foundation, the Heising-Simons Foundation, and the Participating Institutions. SDSS acknowledges support and resources from the Center for High Performance Computing at the University of Utah. The SDSS website is \url{www.sdss5.org}.

The fabrication of the proto-model siderostat was supported by the Korea Astronomy and Space Science Institute under the R\&D program (Project No. 2022-1-860-03) supervised by the Ministry of Science and ICT (MSIT).

SDSS is managed by the Astrophysical Research Consortium for the Participating Institutions of the SDSS Collaboration, including the Carnegie Institution for Science, Chilean National Time Allocation Committee (CNTAC) ratified researchers, the Gotham Participation Group, Harvard University, The Johns Hopkins University, L'Ecole Polytechnique F\'{e}d\'{e}rale de Lausanne (EPFL), Leibniz-Institut f\"{u}r Astrophysik Potsdam (AIP), Max-Planck-Institut f\"{u}r Astronomie (MPIA Heidelberg), Max-Planck-Institut f\"{u}r Extraterrestrische Physik (MPE), Nanjing University, National Astronomical Observatories of China (NAOC), New Mexico State University, The Ohio State University, Pennsylvania State University, Smithsonian Astrophysical Observatory, Space Telescope Science Institute (STScI), the Stellar Astrophysics Participation Group, Universidad Nacional Aut\'{o}noma de M\'{e}xico, University of Arizona, University of Colorado Boulder, University of Illinois at Urbana-Champaign, University of Toronto, University of Utah, University of Virginia, and Yale University.

This manuscript contains the scientific contents previously published in SPIE proceeding at Astronomical Telescopes + Instrumentation 2022\cite{Ahn+22}.

\subsection* {Code and Data Availability} 
The data presented in this paper are publicly available in LVMAGP GitHub repository. The source code of LVMAGP can be found at \url{https://github.com/sdss/lvmagp}.

%%%%% References %%%%%

\bibliography{report}   % bibliography data in report.bib
\bibliographystyle{spiejour}   % makes bibtex use spiejour.bst

%%%%% Biographies of authors %%%%%

\vspace{2ex}\noindent\textbf{Hojae Ahn} is a PhD student at Kyung Hee University, Republic of Korea. He received his BS degree in astronomy and space science from the same university in 2020. His current research interests include astronomical instrumentation (optical design and instrument control software development) and optical observation.

\vspace{2ex}\noindent\textbf{Changgon Kim} is a graduate student at Kyung Hee University, Republic of Korea. He received his BS degree in astronomy and space science from the same university in 2021. His current research interests include astronomical instrumentation, control software, infrared optical design, optical photon simulation, and optical telescopes for space satellites.

\vspace{2ex}\noindent\textbf{Tae-Geun Ji} is a graduate student at the Kyung Hee University, Republic of Korea. He received has BS degrees in astronomy and space science from the KHU in 2016. His current research interests include astronomical instrumentation, software engineering for ground-based optical telescopes.

\vspace{2ex}\noindent\textbf{Soojong Pak} is a professor at the Department of Astronomy and Space Science in Kyung Hee University. He has participated in various astronomical instrumentation and observation projects. His current research projects include the development of instrument control software for the IGRINS, GMTNIRS, and SDSS-V/LVM and optical modules for the MATS and other space satellites.

\vspace{1ex}
\noindent Biographies and photographs of the other authors are not available.

\listoffigures
\listoftables

\end{spacing}
\end{document}